\documentclass[%
reprint, superscriptaddress,
 amsmath,amssymb,
 aps, 
 pra
]{revtex4-2}

\pdfoutput=1
\usepackage[utf8]{inputenc}
\usepackage[english]{babel}
\usepackage[T1]{fontenc}
\usepackage{amsmath}
\usepackage{hyperref}
\usepackage{pdfpages}

\makeatletter
\AtBeginDocument{\let\LS@rot\@undefined}
\makeatother

\usepackage{amsfonts}
\usepackage{physics}
\usepackage{graphicx}
\usepackage{overpic}
\usepackage{color}
\usepackage{comment}
\usepackage{subcaption}

\usepackage{enumitem}
\usepackage{booktabs}
\usepackage{float}

\usepackage{soul}

\usepackage{cancel}
\usepackage{relsize}
\usepackage{cleveref}
\usepackage{mathtools,txfonts,dsfont}
\usepackage{lipsum}

\newcommand{\uulm}{Institute for Complex Quantum Systems, 
  Ulm University,
  Albert-Einstein-Allee 11, 89069 Ulm, Germany}
\newcommand{\unipd}{Dipartimento di Fisica e Astronomia "G. Galilei" \& Padua Quantum Technologies Research Center,
 Universit{\`a} degli Studi di Padova, 35131 Padova, Italy}
\newcommand{\pdinfn}{INFN, Sezione di Padova, via Marzolo 8, 35131
  Padova, Italy}
\newcommand{\juelich}{Peter Grünberg Institute -- Quantum Control (PGI-8), Forschungszentrum J\"ulich GmbH, 52425 J\"ulich, Germany}
\newcommand{\koeln}{Institute for Theoretical Physics, University of Cologne, 50937 K\"oln, Germany}
\newcommand{\bologna}{Dipartimento di Fisica e Astronomia, Università di Bologna, 40127 Bologna, Italy}
\newcommand{\tuwien}{Atominstitut, Technische Universit{\"a}t Wien, Stadionallee 2, 1020 Wien, Austria}

\begin{document}


\title{The Role of Bases in Quantum Optimal Control}

\author{Alice Pagano}
\affiliation{\uulm}
\affiliation{\unipd}
\affiliation{\pdinfn}

\author{Matthias M Müller}
\affiliation{\juelich}

\author{Tommaso Calarco}
\affiliation{\juelich}
\affiliation{\koeln}
\affiliation{\bologna}

\author{Simone Montangero}
\affiliation{\uulm}
\affiliation{\unipd}
\affiliation{\pdinfn}

\author{Phila Rembold}
\affiliation{\unipd}
\affiliation{\koeln}
\affiliation{\tuwien}

\date{\today}

\begin{abstract}
Quantum Optimal Control (QOC) supports the advance of quantum technologies by tackling its problems at the pulse level: Numerical approaches iteratively work towards a given target by parametrising the applied time-dependent fields with a finite set of variables. The effectiveness of the resulting optimisation depends on the complexity of the problem and the number of variables. We consider different parametrisations in terms of basis functions, asking whether the choice of the applied basis affects the quality of the optimisation. Furthermore, we consider strategies to choose the most suitable basis. 
For the comparison, we test three different randomisable bases -- introducing the sinc and sigmoid bases as alternatives to the Fourier basis -- on QOC problems of varying complexity. For each problem, the basis-specific convergence rates result in a unique ranking. Especially for expensive evaluations, e.g., in closed-loop, a potential speed-up by a factor of up to 10 may be crucial for the optimisation's feasibility. We conclude that a problem-dependent basis choice is an influential factor for QOC efficiency and provide advice for its approach. 
\end{abstract}

\maketitle


\section{Introduction}
\label{sec:introduction}

Quantum optimal control theory resulted in the development of a family of algorithms that iteratively adjust time-dependent control fields to improve a given quantum process~\cite{Koch2022, Glaser2015, Brif2010, Rembold2020, Muller2022}. The quality of said process is quantified by a figure of merit (FoM) which is minimised (or maximised). When the full classical description of the quantum system is available, the most cost-efficient way to optimise them is typically via a numerical simulation of this model, or in other words with an open-loop approach~\cite{Omran2019,Pagano2022}.
However, when the system is expensive to characterise in its entirety -- including transfer functions, noise, environmental factors, etc. -- a closed-loop approach~\cite{Rosi2013, vanFrank2016, Heck2018, Cerfontaine2020, werninghaus_leakage_2021, oshnik_robust_2022, Vetter2024} is more effective. This type of optimisation relies on a direct connection between the experiment and the algorithm. In most optimisation algorithms the control pulses are expanded in terms of basis functions. Certain methods such as (d)CRAB~\cite{Caneva2011,Rach2015} or GOAT~\cite{goat} typically use smooth basis functions, such as the (random) Fourier basis. 
Other methods like GRAPE or Krotov split the pulse into piecewise constant segments~\cite{KhanejaGRAPE}. 
In the context of this work we interpret them as elements of a fixed-width square-pulse basis. 
While the basis does not have to be orthogonal, it must cover a suitable area of the accessible 
function space in order not to impose a too tight constraint. Many bases meet these criteria. 
For open-loop optimisations, where the gradients of the FoM are available, the piecewise-constant basis~\footnote{see Supplementary Material I} is very popular as it mirrors the most common modelling techniques for quantum dynamics~\cite{Konnov1999}. 
The information provided by an accessible gradient allows gradient-based techniques to efficiently explore large candidate spaces. 
However, it has been shown that even for gradient-based methods employing other bases may increase accuracy, flexibility, efficiency, and smoothness~\cite{goat, group,motzoi_optimal_2011}.
The introduction of a truncated basis can also be studied from an information-theoretical perspective by comparing the information content of the control to the number of degrees of freedom required by the FoM (dimensionality of the problem). From such considerations it follows that the truncated basis has to be large enough to match the dimensionality of the problem~\cite{Lloyd2014,Caneva2014,Muller2022SR}.
Nevertheless, the basis is often chosen according to the default associated with a given QOC algorithm 
with the common attitude that any large enough set should equivalently be able to find the same solutions. 
Below, we show how such an attitude can adversely affect the efficiency of the optimisation.

\emph{Motivation --}
Two main arguments support the careful consideration of the basis:
First, the number of basis elements required to reconstruct a known solution clearly depends on the chosen basis: It is simple to assemble a fast oscillating pulse with Fourier elements but complex with the piecewise-constant basis. For some problems, all local solutions share certain traits. As a result, bases that can reproduce the relevant pulse properties will be more likely to converge to a good local optimum. Examples are provided in Ref.~\cite{Jensen2021,li_optimal_2023}, where the authors show that the found solutions always consist of linear and bang-bang segments, which require high complexity when decomposed via the Fourier basis. 
A further illustration of the significance of a problem-tailored basis is provided in the context of noise characterisation~\cite{chalermpusitarak_frame-based_2021}, where it determines how well the system properties can be reconstructed. 
Second, within a basis, each element has a distinct impact on the system. 
Considering a non-contributing basis element whose adaption results in no change of the FoM, will result in a slower convergence. This could, e.g., be the case for a frequency component that exceeds the time scale of the system or, in some cases, a basis element that does not change the integral of the pulse. 
Lastly, a basis may be chosen deliberately to encode control limitations such as limited rise times or bandwidth restrictions~\cite{oshnik_robust_2022, goat}. 

\emph{Approach --}
To systematically explore the impact of the basis choice on the convergence of the optimisation, we consider two well-explored QOC problems: The state transfer in an Ising chain controlling only the coupling between spins~\cite{Rach2015}, and the optimisation of a qubit NOT gate in the presence of a third, slightly detuned energy level~\cite{Motzoi2009}. For each example, we compare three bases chosen for their distinct properties. The optimisations are executed using the dCRAB (dressed Chopped RAndom Basis~\cite{Rach2015, Muller2022}) formalism in a gradient-free setting to show their applicability in situations with limited access to information about the system and some are reproduced using automatic differentiation~\footnote{see Supplementary Material III}. The majority of optimisations was implemented using the QuOCS software suite~\cite{Rossignolo2023}. 
We find that the convergence rate of the optimisation is highly dependent on the employed basis and that their ranking differs for each problem. The reasons for that become clearer when analysing the average solutions. We conclude that a problem-dependent basis choice is an influential factor for QOC efficiency. Hopefully, our results will lead to more efficient applications of QOC in the future.

\emph{Outline --}
Sec.~\ref{sec:bases} starts with a description of the different bases. In Sec.~\ref{sec:examples} we examine two 
distinct physical models and assess how the choice of basis affects the associated optimisation tasks. From those examples we formulate some advice for picking a basis in Sec.~\ref{sec:advice}. Lastly, the results are summarised in Sec.~\ref{sec:conclusion}, where we give an outlook.


\section{Bases}
\label{sec:bases}

In this work, pulses are constructed according to the dCRAB formalism~\cite{Muller2022}. Here, a given number $N_s$ of basis vectors is optimised at a time. This process is referred to as a superiteration. After it converges, a new basis vector is constructed from randomly assembled basis elements each characterised by a superparameter $s$. This basis vector is added to the previously converged solution $c_\text{fin}^{J-1}(t)$ and optimised in the next superiteration. Each basis element $f(s, \vec{A}; t)$ is defined by the corresponding optimisation parameter(s) $\vec{A}$. The full pulse $c_i^J(t)$ at iteration $i$ within the $J^\text{th}$ superiteration is written as
\begin{equation}
\begin{split}
    c_i^J(t) &= c_\text{fin}^{J-1}(t) + u_i^J(t)\\
    &= \sum_{j=1}^{J-1} u_\text{fin}^j (t) + u_i^J(t),
\end{split}
\end{equation}
where $u_i^J(t)$ is the pulse optimised in each superiteration with $N_s$ superparameters, given by:
\begin{equation}
    u_i^J(t) = \sum_{n=1}^{N_s} f(s_n^J, \vec{A}_{i}; t)\; \Bigl[ + f_0 (A_{0i}; t) \Bigr].
\end{equation}
The last term $f_0$ refers to a special basis element that is optimised in every superiteration, which is only relevant for the basis in Sec.~\ref{sec:sigmoid}.

The following bases have been chosen as they fit in the dCRAB framework and represent a diverse set of properties. For that reason the popular piece-wise constant basis was replaced by the sigmoid basis in this work~\footnote{see Supplementary Material I}.

\subsection{Fourier}
\begin{figure}[h!]
    \centering
    \includegraphics[width=\linewidth]{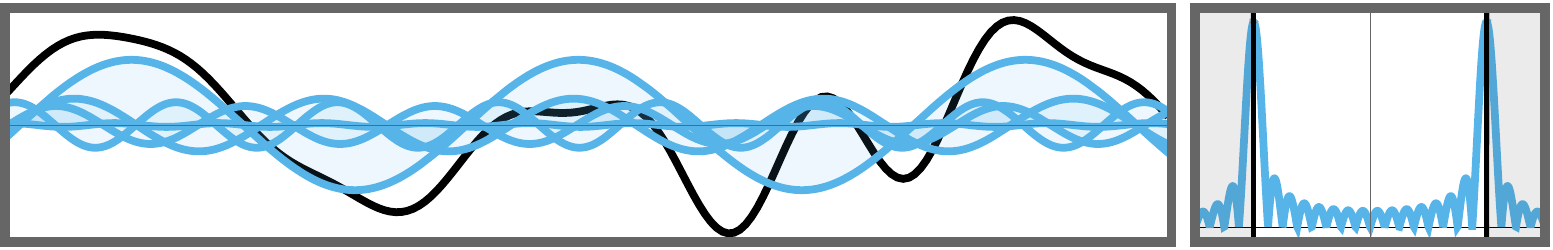}
    \caption{Fourier basis. Left: An example pulse (black) decomposed with Fourier elements (blue). Right: Spectrum of a basis element with $s=\omega_\text{max}$ indicated by vertical lines. The cut-off at the start and end time of the pulse leads to the spillage to higher frequencies.}
    \label{fig:fourier}
\end{figure}

The Fourier basis~\cite{Caneva2011} is composed of frequency elements. The randomised superparameter corresponds to the frequency $s$. For each superparameter, there are two optimisation parameters $A=[A^1,A^2]$. Thus, the number of optimisation parameters is $N_\text{opt}=2 N_s$ and a basis element can be written as:
\begin{equation}
    f(s,[A^1,A^2];t)=A^1 \sin(s t) + A^2 \cos(s t).
\end{equation}
Frequency limits are implemented via a corresponding bandwidth window $0<s\le \omega_\text{max}$ to choose the superparameters from. The superparameters are picked at random from equally distributed bins in the frequency range (see Ref.~\cite{Rembold2020} for a detailed explanation).
An example of pulse constructed by Fourier elements is depicted in Fig.~\ref{fig:fourier}.

\subsection{Sinc}

\begin{figure}[h]
    \centering
    \includegraphics[width=\linewidth]{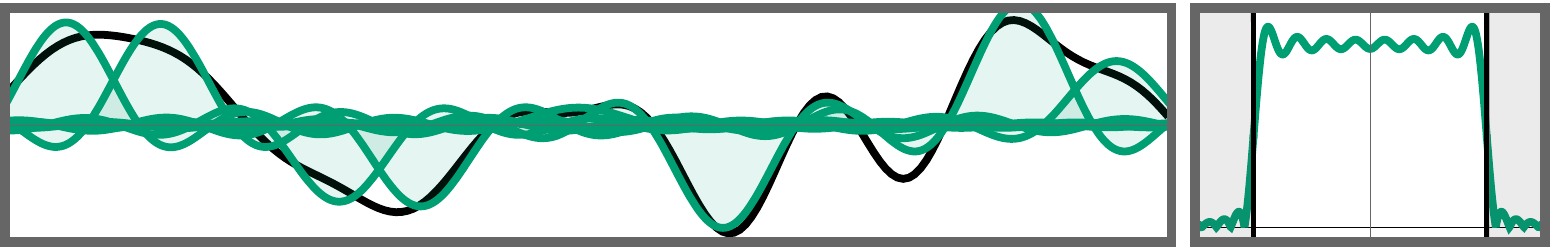}
    \caption{Sinc basis. Left: The same example pulse as in Fig.~\ref{fig:fourier} (black) decomposed with sinc basis elements (green). Right: Spectrum of a basis element with $s=0.5T$. The vertical lines indicate $\pm\omega_\text{max}$. The cut-off at the start and end time of the pulse leads to the spillage to higher frequencies.}
    \label{fig:sinc}
\end{figure}

The sinc basis is composed of wave packets with a fixed bandwidth $\omega_\mathrm{max}$. The randomised superparameter corresponds to the center time of the wave packet. The amplitude $A$ of each wave packet is optimised. The basis element $f$ is:
\begin{equation}
    f(s,A;t) = A  \text{sinc}(\omega_\mathrm{max} (t-s)).
\end{equation}
The number of optimisation parameters is $N_\text{opt}= N_s$ and the frequency is limited by $\omega_\text{max}$ as the spectrum of a sinc function is a top-hat function of width $\omega_\text{max}$, as shown in Fig.~\ref{fig:sinc}.

\subsection{Sigmoid}\label{sec:sigmoid}
\begin{figure}[h]
    \centering
    \includegraphics[width=\linewidth]{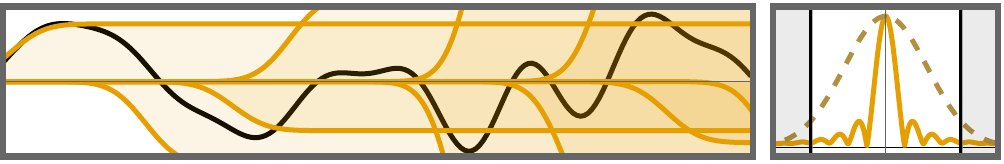}
    \caption{Sigmoid basis. Left: The same example pulse as in Fig.~\ref{fig:fourier} (black) decomposed with sigmoid basis elements (yellow). Right: Spectrum of a basis element with $s=0.5T$ cut off at beginning and end of the pulse. The vertical lines indicate $\omega_\text{max}$. The cut-off at the start and end time of the pulse leads to the spillage to higher frequencies. However, sigmoid spectra are on average  The dashed line indicates the maximum frequency envelope for an ideal sigmoid pulse (see supplementary material).}
    \label{fig:sigmoid}
\end{figure}

The sigmoid basis~\cite{rembold_quantum_2022, oshnik_robust_2022} is composed of smooth time steps, see Fig.~\ref{fig:sigmoid}. The randomised superparameter corresponds to the time of the step. The amplitude $A$ of each step is optimised together with an initial amplitude $A_0$ at time zero. The number of optimisation parameters is $N_\text{opt}=N_s + 1$.  Each basis element can be written as:
\begin{equation}
\begin{split}
    f(s,A;t) &= \frac{A}{2}\left(1+\text{erf}\left(\frac{t-s}{\sqrt{2}\sigma}\right)\right),\\
    f_0(A_0;t) &= f(0,A_0;t).
\end{split}
\end{equation}
The frequency is limited by setting the width of each step $\sigma$:   
\begin{equation}
    \sigma = \sqrt{\frac{-2 \ln \epsilon_\text{cut}}{\omega_\text{max}^2}}.
\end{equation}
The envelope of the pulse spectrum is given by a Gaussian. To limit the frequency in a comparable manner with respect to the other bases, we require the Gaussian envelope to be reduced to $\epsilon_\text{cut}=0.2$ at $\omega_\text{max}$. See supplementary material for further discussion on the bandwidth limitations.

\section{Test Problems}
\label{sec:examples}

The examples were picked to represent different QOC problems which have previously been shown to be optimisable~\cite{Rach2015,Motzoi2009}. Hence, they represent a good point of comparison between different bases. We investigate two factors to evaluate the quality of each basis for a given problem. First, the \textbf{convergence probability} $P_c$, i.e., the percentage of optimisations which converges below a given threshold $F_\text{conv}$ within a maximum number of function evaluations. 
Second, the \textbf{convergence period} $\nu_c$, i.e., the median of function evaluations needed for the converging optimisations to reach said threshold.

\begin{figure*}[t]
   \begin{center}
     \begin{minipage}{0.325\textwidth}
     \begin{overpic}[width=1.05 \textwidth,unit=1mm]{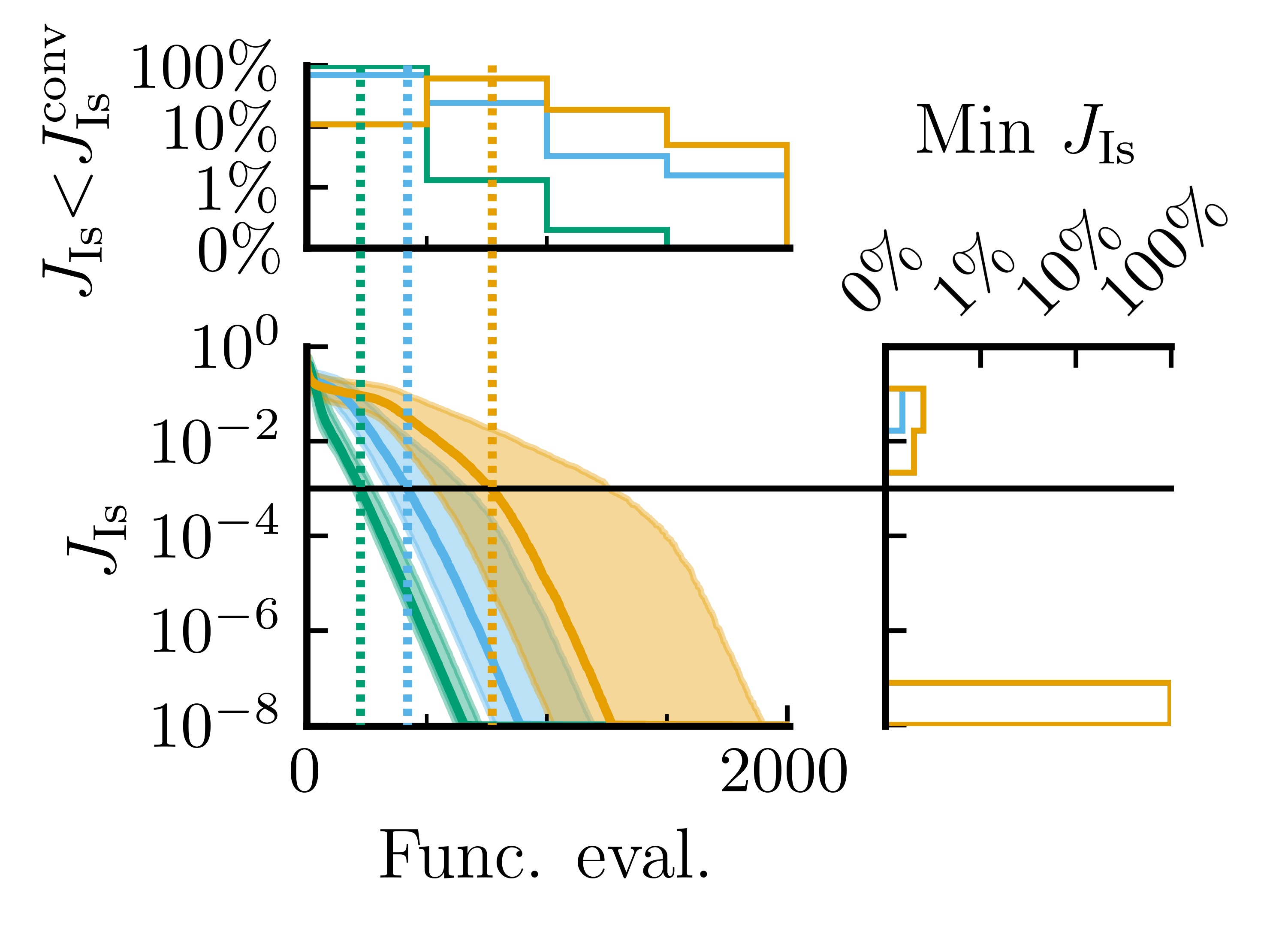}
        \put(-1,78){\textbf{(a)} 2-qubit}
      \end{overpic}
    \end{minipage}
     \begin{minipage}{0.325\textwidth}
        \begin{overpic}[width=1.05 \textwidth,unit=1mm]{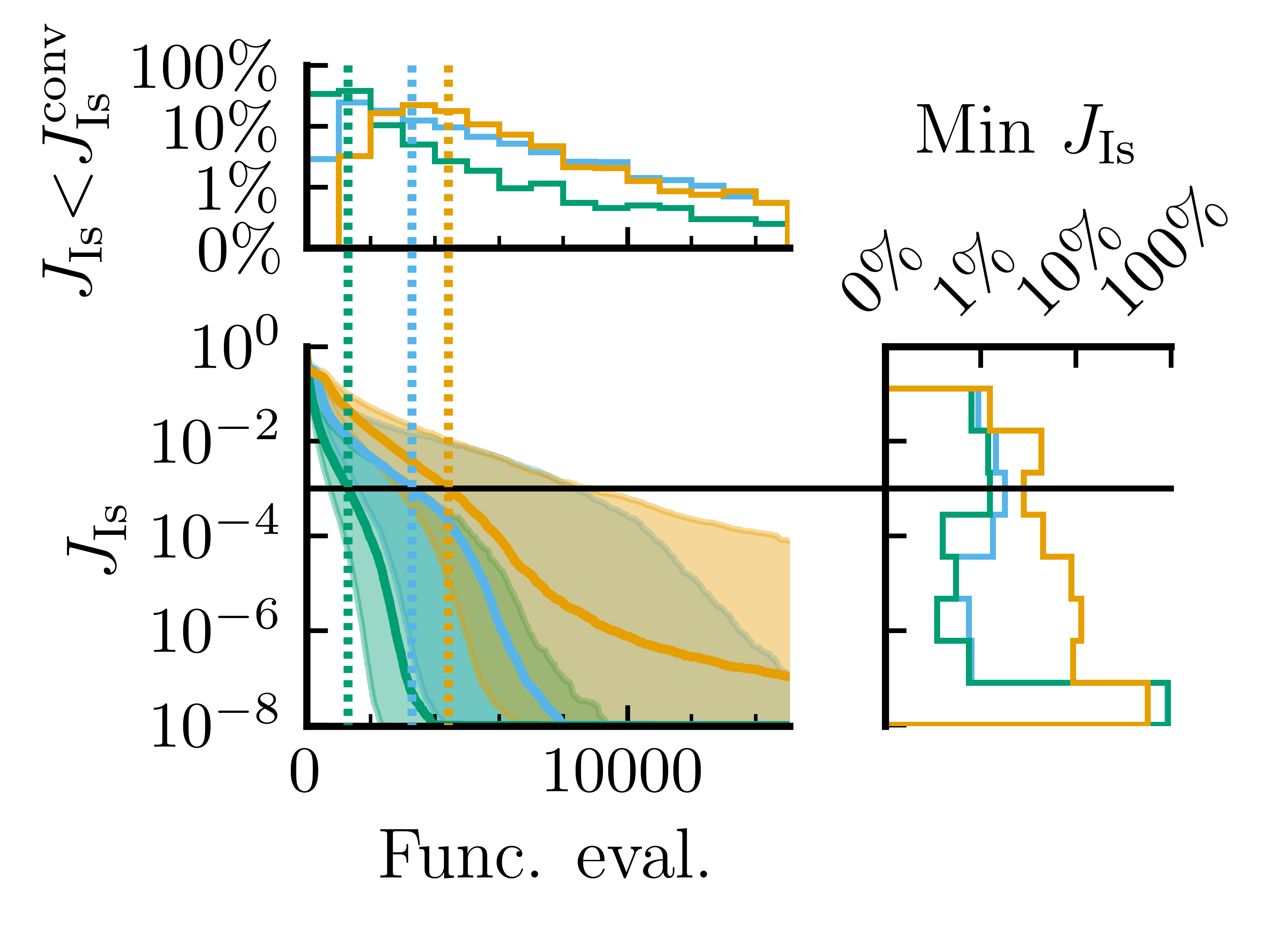}
        \put(-1,78){\textbf{(b)} 3-qubit}
        \end{overpic}
    \end{minipage}
     \begin{minipage}{0.325\textwidth}
        \begin{overpic}[width=1.05 \textwidth,unit=1mm]{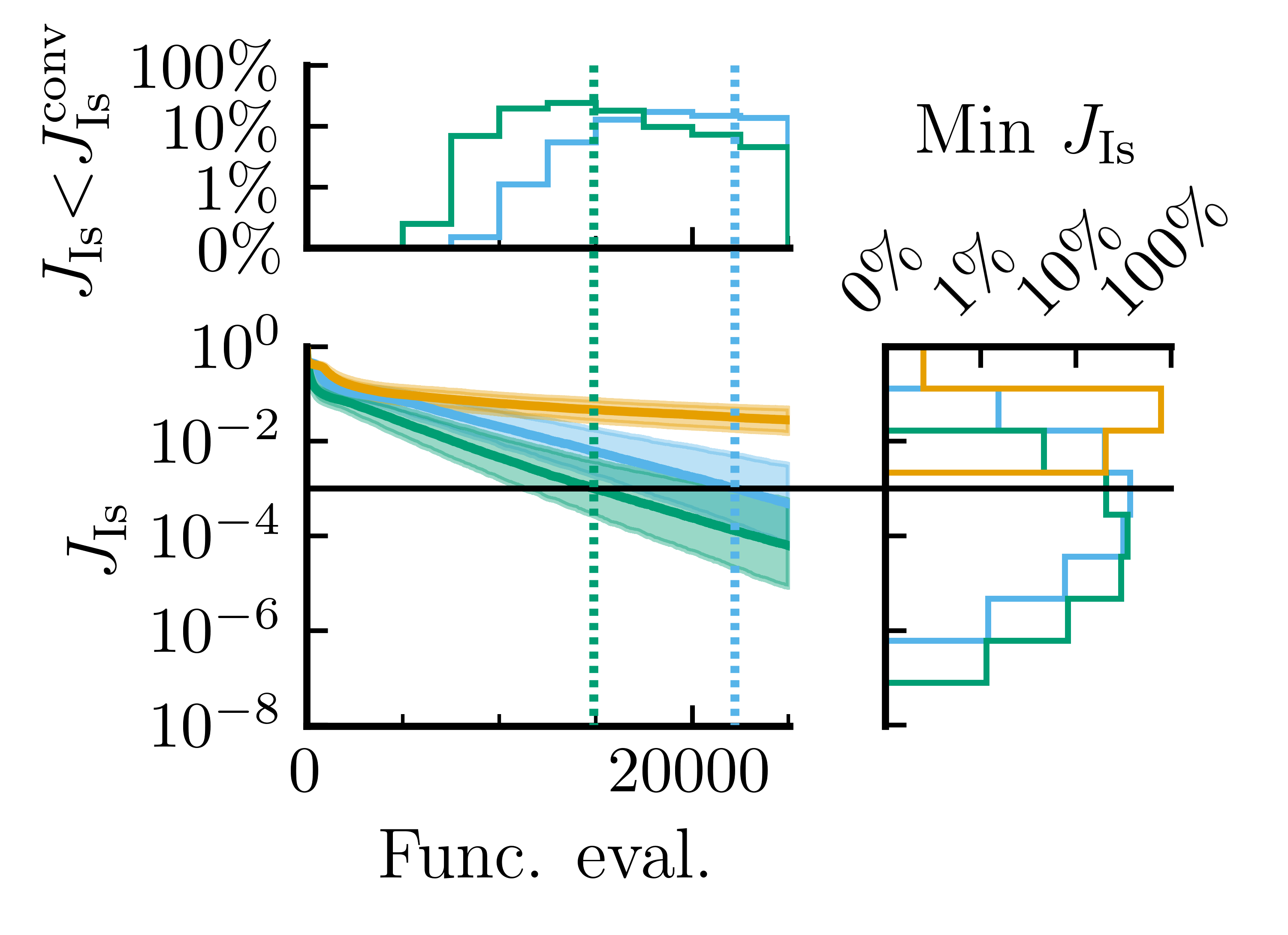}
        \put(-1,78){\textbf{(c)} 4-qubit}
        \end{overpic}
    \end{minipage}
    \centering
     \begin{minipage}{0.325\textwidth}
        \begin{overpic}[width=1 \textwidth,unit=1mm]{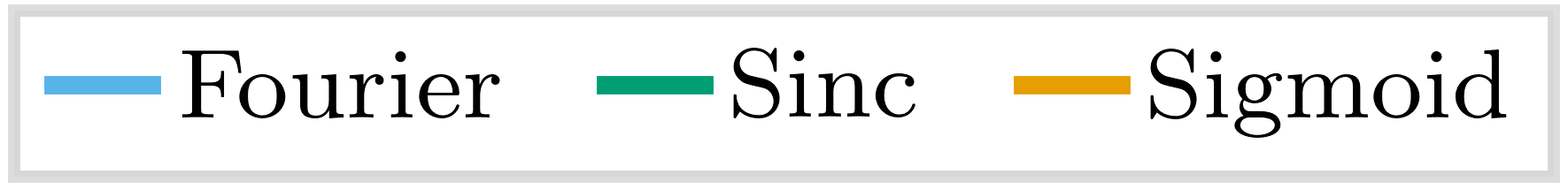}
        \end{overpic}
    \end{minipage}
    \caption{Median convergence of the Ising model problem for 2, 3, and 4 qubits with different bases. The main plots show the median convergence trace with a shaded area representing the 16th and 84th percentiles, i.e. corresponding to $68\%$ of the optimisations. The black horizontal line represents the threshold considered for convergence at $J_\text{Is}^\text{conv} = 10^{-3}$. The dotted lines show the convergence period $\nu_c$ given in Table~\ref{tab:Ising}. The histograms above represent the percentage of optimisations that have dropped below that threshold within the bin's number of function evaluations. Similarly, the histograms on the right show the final values of the cost $J_\text{Is}$. All optimisations were stopped at $10^{-8}$ or at $25000$ iterations.}
    \label{fig:ising_traces}
  \end{center}
\end{figure*}

In order for the optimisations to be comparable we set the maximum number of function evaluations to $25\,000$, use the same number of optimisation parameters $N_\text{opt}$, the same pulse time $T$, and the same maximum frequency $\omega_\text{max}$ for each basis. The hyperparameters, including $T$ and $\omega_\text{max}$, are chosen to give the best average results for convergence probability. Exemplary results for alternative hyperparameter settings are provided in the supplementary material~\footnote{see Supplementary Material II} and show that they do not affect the rankings between the bases. 
Every optimisation is repeated $2,000$ times with different random superparameter $s$ to obtain the average convergence plot.
Finally, we investigate common features of the best solutions.

\subsection{Random Ising Chain}
\label{sec:ising}

The first example is taken in direct analogy to Ref.~\cite{Rach2015}. Here, the authors showed how an optimisation with the Fourier basis converges depending on the number of optimisation parameters used. Hence, it is an ideal testbed for a comparison with other bases.
We consider a chain composed of $N$ spins controlled solely via the global interactions between nearest neighbors with the following Hamiltonian:
\begin{equation}
    H_\text{Is}/\hbar=\sum_{n=1}^N \alpha_n \sigma_n^x+\beta_n \sigma_n^z+c(t) \sum_{n=1}^{N-1} \sigma_n^z \sigma_{n+1}^z.
    \label{eq:Hamiltonian_Ising}
\end{equation}
Each optimisation run is performed with random coefficients $\alpha_i,\,\beta_i \in [0,1]$ and random initial and target states, $\ket{\psi_0}$ and $\ket{\psi_\text{t}}$. The FoM is given by the final state infidelity
\begin{equation}
    J_\text{Is}=1-|\bra{\psi_\text{t}}U(T)\ket{\psi_0}|^2,
\end{equation}
where $U(T)$ is the time propagator corresponding to Eq~\eqref{eq:Hamiltonian_Ising}. The problem becomes harder to solve as the number of qubits $N$ goes up. We have investigated the convergence for $N=\{2,3,4\}$. The convergence probabilities $P_c$, convergence period $\nu_c$, and hyperparameters for the different bases are summarised in Table~\ref{tab:Ising}.

\begin{table}[h]
    \centering
    \begin{tabular}{c c |c|c|c}
         \multicolumn{2}{c|}{$N$} & 2 & 3 & 4  \\
         \hline
         \hline
         \multicolumn{2}{c|}{$T$} & 20 & 20 & 75  \\
         \multicolumn{2}{c|}{$\omega_\text{max}$} & $2 \pi 20$ & $2 \pi 20$ & $2 \pi 75$  \\
         \multicolumn{2}{c|}{ $N_\text{opt}$} & 12 & 16 & 16  \\
         \hline
         \hline
         Fourier & $P_c$ & \textbf{99.75\%} & \textbf{96.75\%} & \textbf{65.05\%}  \\
                & $\nu_c$ & 420 & 3273 & 22217 \\
         \hline
         sinc & $P_c$& \textbf{99.8\%} & \textbf{97.25\%} & \textbf{90.1\%}  \\
                & $\nu_c$ & 224 & 1291 & 14898  \\
         \hline
         sigmoid & $P_c$& \textbf{99.45\%} & \textbf{93.5\%} & \textbf{0\%}  \\
                & $\nu_c$ & 771 & 4410 & $-$  \\
    \end{tabular}
    \caption{Ising model optimisations. The hyperparameters for 2, 3, and 4 qubits are accompanied by the convergence probabilities $P_c$, and convergence periods $\nu_c$ for the three tested bases shown in Fig.~\ref{fig:ising_traces}.}
    \label{tab:Ising}
\end{table}

\begin{figure}[b]
   \begin{center}
     \begin{minipage}{0.98\linewidth}
        \includegraphics[width=1\textwidth]{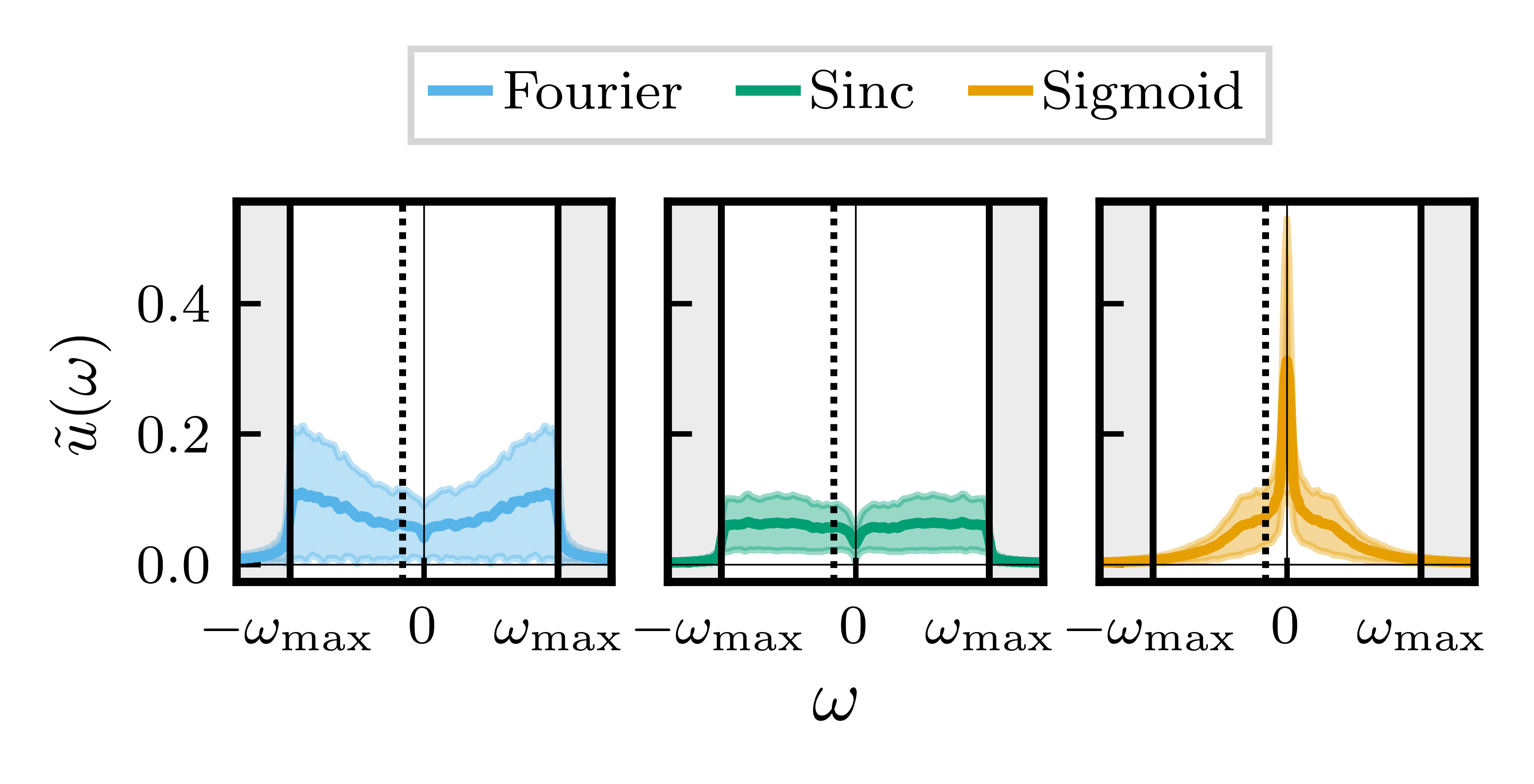}
    \end{minipage}
    \caption{Solutions of the 3-qubit Ising problem: Average spectra for the converged solutions ($J_\text{Is}<10^{-3}$) with different bases. The vertical lines indicate $\omega_{\mathrm{max}}$.}
    \label{fig:ising_solution}
  \end{center}
\end{figure}

In Fig.~\ref{fig:ising_traces}, we show the median convergence traces of the optimisations. The plots identify a clear winner in terms of convergence periods for all $N$: The sinc basis is fastest, followed by Fourier and eventually sigmoid. For two and three qubits most optimisations reach the threshold $J_\text{Is} < 10^{-3}$. Here, the sinc basis is approximately twice as fast as Fourier and three times as fast as sigmoid. For four qubits more iterations would have been necessary to see convergence for the sigmoid basis while Fourier and sigmoid converge for a majority of optimisations.
The average spectra of the converged pulses in Fig.~\ref{fig:ising_solution} show the big difference between the bases: While the Fourier and sinc basis result in broad spectra with a reduced amplitude at frequency zero, the sigmoid basis remains strongly focused on zero. This result is not surprising as the complexity of building a pulse with a strong frequency component solely with step functions becomes higher for higher frequencies. The absence of high frequency components may be responsible for the sigmoid basis' poor performance for this test problem.

\subsection{Single Qubit Gate on a Qutrit}
\label{sec:qubit_gate}

Most qubit systems are constructed by isolating two levels from a many-level Hamiltonian. One example can be found in NV (nitrogen-vacancy) centers in diamond~\cite{doherty_nitrogen-vacancy_2013}, where the native three-level system is reduced to two, even if at low magnetic field one has to take all three levels into account~\cite{Vetter2022}. 
Another popular representative are superconducting circuits which can be approximated as qubits due to their anharmonicity $\Delta$. The anharmonicity describes the degree to which the system deviates from a simple harmonic oscillator and hence, how far detuned the $\ket{1}\leftrightarrow\ket{2}$ transition is from the $\ket{0}\leftrightarrow\ket{1}$ transition. Without it, it would not be possible to address the levels separately. Still, even with anharmonicity the transitions are typically close which can lead to leakage and errors in single qubit gates. A number of strategies exists to avoid this issue~\cite{krantz_quantum_2019, caneva_optimal_2009, Safaei2009} the most famous of which is likely DRAG~\cite{Motzoi2009}. This analytic strategy provides controls that prevent a majority of leakage and is explained in further detail in the supplementary material~\footnote{see Supplementary Material V}. In zeroth order it reduces the frequency component at $\Delta$ majorly responsible for the unwanted coupling to zero. However, QOC methods -- specifically the gradient-based method GRAPE -- have been shown to outperform it with few parameters~\cite{Motzoi2009}. A piecewise-constant basis has even been applied recently to optimise a DRAG-based NOT gate in closed-loop using superconducting qubits~\cite{werninghaus_leakage_2021}.

\begin{figure}[t]
   \begin{center}
     \begin{minipage}{0.99\linewidth}
        \includegraphics[width=1\textwidth]{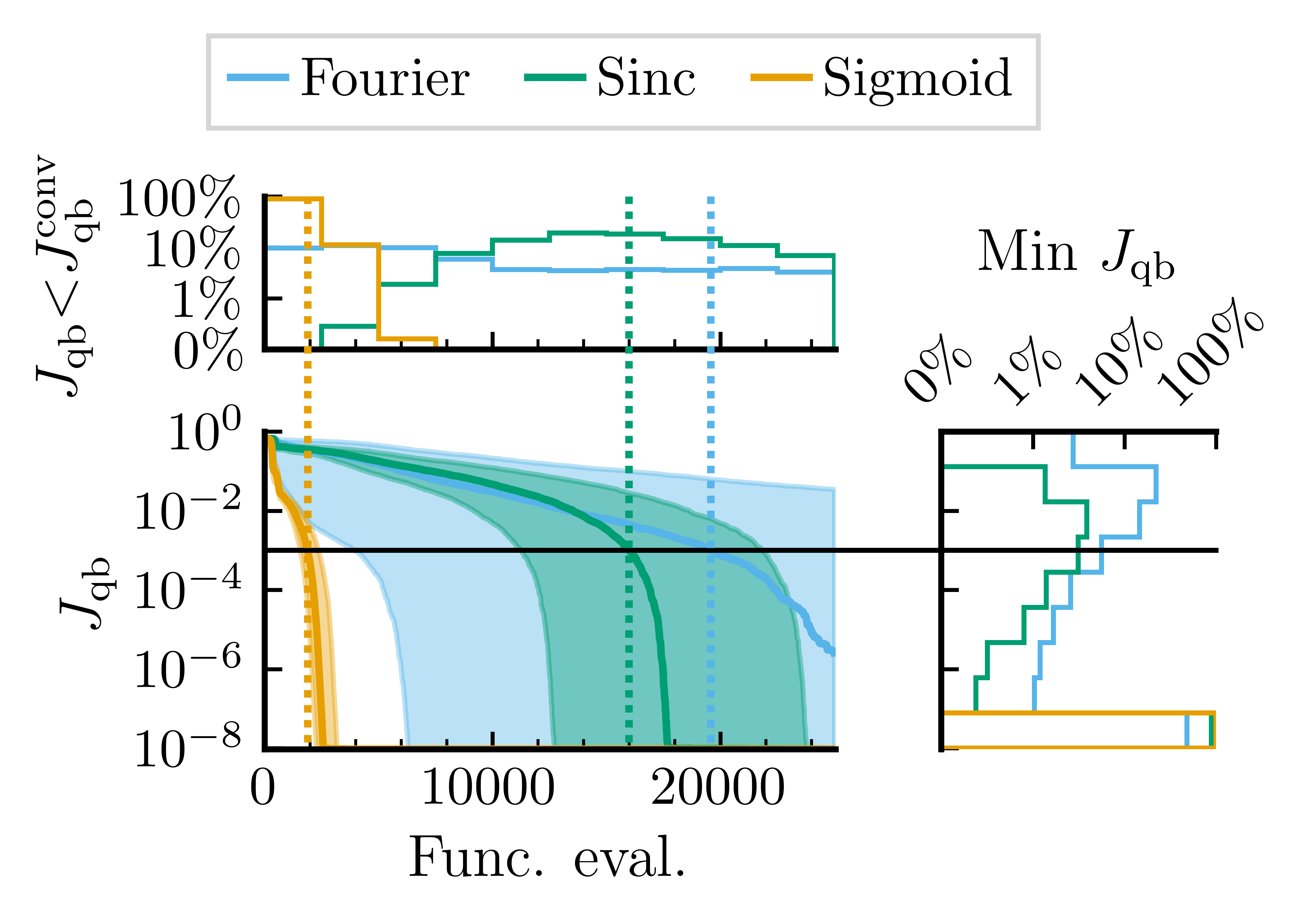}
    \end{minipage}
    \caption{
    Convergence with different bases optimising a qubit NOT gate. The main plot shows the median convergence trace with a shaded area representing the 16th and 84th percentiles.
    The histogram on the right shows the distribution over the final FoM, either at $10^{-8}$ or after 25000 iterations. All optimisations that reach the threshold $J_\text{qb}^\text{conv} = 10^{-3}$, represented by the black line, are considered converged. The percentage of optimisations that are converged within the number of iterations are illustrated in the histogram above.
    }
    \label{fig:qubit_traces}
  \end{center}
\end{figure}

After the application of the rotating wave approximation the three-level system is described by the following Hamiltonian:
\begin{equation}
\begin{split}
    H_\text{qb}/\hbar&= \Delta \ket{2}\bra{2} + \delta(t) \ket{1}\bra{1}\\
    &+\sum_{n=1,2}\frac{\sqrt{n}}{2}\left[c_x(t) \sigma_{n-1,n}^x+c_y(t)\sigma_{n-1,n}^y\right],
\end{split}
\label{eq:Hamiltonian_qubit}
\end{equation}
where $c_x$ and $c_y$ are the controls on $\sigma_x^{n,m}=\ket{n}\bra{m}+\ket{m}\bra{n}$, $\sigma_y^{n,m}=i(\ket{m}\bra{n}-\ket{n}\bra{m})$, and $\delta$ represents the detuning of the controls. In the following optimisation, $\delta=0$ and only the $x$- and $y$-components of the drive are considered, since a gradual phase change between the components also leads to an effective detuning. The factors $\sqrt{n}$ represent the couplings between the different levels. To calculate the qubit gate fidelity we average over the evolution of all qubit input states $j_\text{in}=\{ \pm x, \pm y, \pm z\}$ lying on the axes of the Bloch sphere. The gate infidelity is then given by~\cite{Motzoi2009}
\begin{equation}
\begin{split}
    J_\text{qb}=1-\frac{1}{6} \sum_{j=j_\text{in}} \left|\bra{j} U_{\text {t }}^{\dagger} U(T)\ket{j}  \right|^2,
\end{split}
\end{equation}
where $U(T)$ is the propagator corresponding to Eq~\eqref{eq:Hamiltonian_qubit} at final time $T$ and $U_\text{t}$ is the target gate defined as
\begin{equation}
    U_\text{t} = \left(\begin{tabular}{c c c}
        0 & 1 & 0 \\
        1 & 0 & 0 \\
        0 & 0 & 1
    \end{tabular}\right).
\end{equation}

\begin{table}[b]
    \centering
    \begin{tabular}{c|c|c}
         Basis & $P_c$ & $\nu_c$  \\
         \hline
         \hline
         Fourier & \textbf{57.64\%} & 19567  \\
         \hline
         sinc & \textbf{93.44\%} & 15987   \\
         \hline
         sigmoid & \textbf{99.96\%} & 1880  \\
    \end{tabular}
    \caption{Convergence probabilities and speeds for the NOT gate using hyperparameters $T=12.5/|\Delta|$, $N_\text{opt}=6$, and $\omega_\text{max}=12.5\pi/|\Delta|$ with $\Delta=-400 \cdot 2\pi\,$MHz chosen to produce efficient optimisations.}
    \label{tab:qubit}
\end{table}

Table~\ref{tab:qubit} summarises the optimisation results from different bases shown in Fig.~\ref{fig:qubit_traces}. The convergence traces indicate that this problem is most efficiently solved with the sigmoid basis. Indeed, the corresponding optimisation is approximately 10 times faster than the other two. The good results are not unexpected as the sigmoid basis can be seen as a dCRAB version of the piecewise-constant basis which has been shown to perform well for this problem~\cite{Motzoi2009,werninghaus_leakage_2021}. More details on their equivalence can be found in the supplementary material~\footnote{see Supplementary Material I}. While the convergence period is similar between the sinc and Fourier basis, the sample variance between Fourier optimisations is large in comparison to the sinc basis. Practically, that means one is more likely to find a good result with the Fourier basis after few iterations than with the sinc basis. However, the probability is also higher to not find a solution at all. This circumstance is illustrated by the top histogram in Fig.~\ref{fig:qubit_traces} and the large spread of the shaded blue area.

\begin{figure}[t]
   \begin{center}
     \begin{minipage}{0.99\linewidth}
     \begin{overpic}[width=1 \columnwidth,unit=1mm]{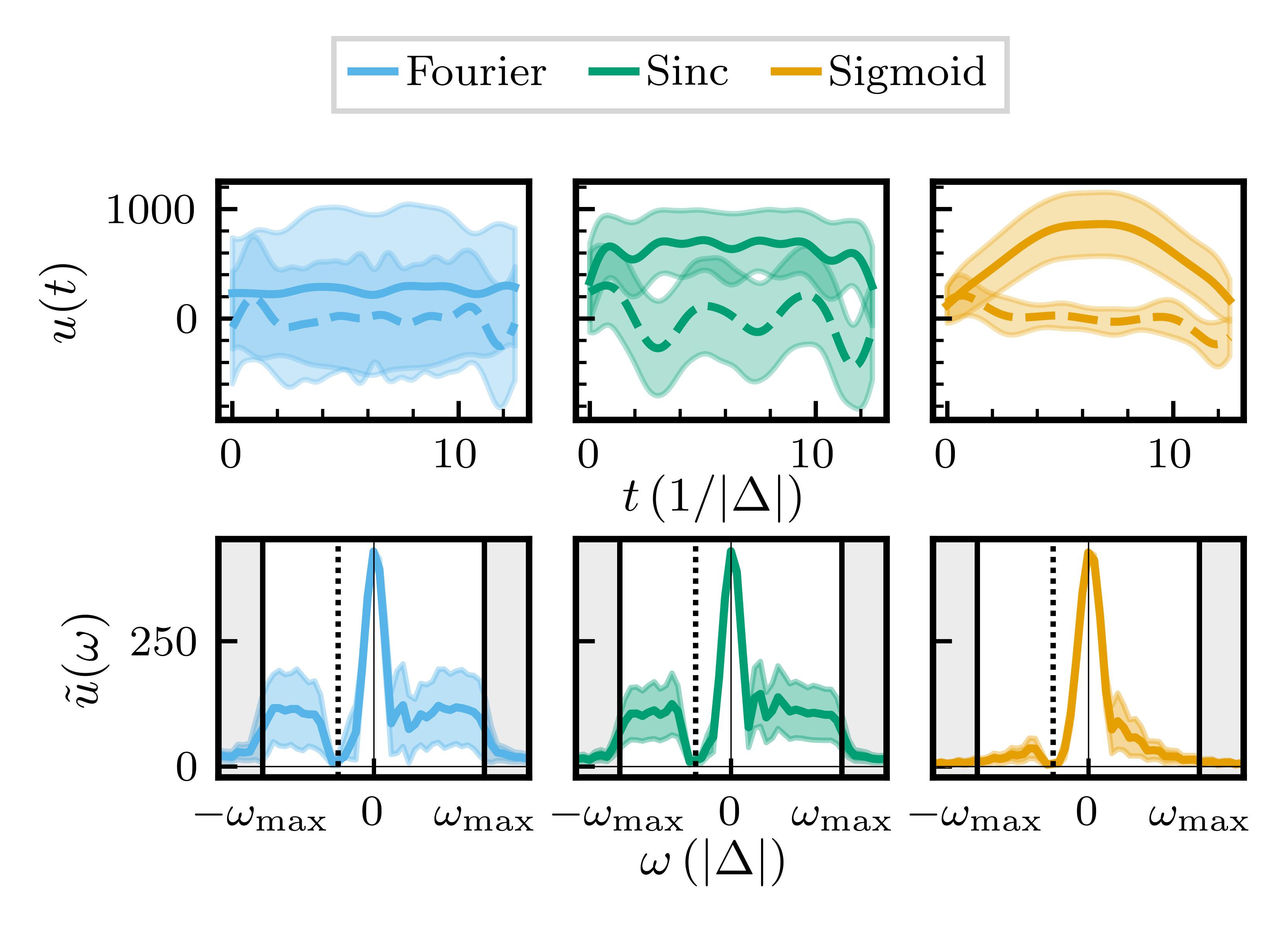}
        \put(0,68){\textbf{(a)}}
        \put(0,38){\textbf{(b)}}
      \end{overpic}
    \end{minipage}
    \caption{\textbf{(a)} Average pulse for the converged solution for different bases. The X and Y pulses are the solid and dashed lines respectively. \textbf{(b)} Corresponding average spectra. The solid vertical lines indicate $\omega_{\mathrm{max}}$ and the dashed line $\Delta$.
    }
    \label{fig:qubit_solution}
  \end{center}
\end{figure}

The differences between these results can be explained with the average solutions displayed in Fig.~\ref{fig:qubit_solution} considering the properties a good solution should have. 
A pulse that flips a qubit in presence of a third level without leakage requires two things: an integral equal to $\pi$ and a hole in its spectrum corresponding to the transition to the third level. The average spectra of the converged pulses in Fig.~\ref{fig:qubit_solution}b show exactly that for all bases. Further information is revealed when comparing the time-dependent pulse shapes to DRAG~\cite{Motzoi2009}. This analytical strategy for finding leakage-free pulses~\footnote{see Supplementary Material V} typically starts with a symmetric $x$-component of the pulse, while the $y$-component is antisymmetric following the $x$-component's derivative. Such behaviour is clearly mirrored in the average sigmoid pulse in Fig.~\ref{fig:qubit_solution}a, but less so for the other two bases. After all, the sigmoid basis only requires one element to build a low-bandwidth, high integral $x$-pulse, and only two for a low-frequency, antisymmetric $y$-pulse. For both Fourier and sinc basis such a construction is more complex. Especially the sinc basis has a naturally wide spectrum, leading to large leakage unless it is actively counteracted, which explains the slow convergence.

\section{Discussion}\label{sec:advice}
The set of existing QOC methods unites different approaches: some gradient-free, some gradient-based, some including an inherent modelling technique, and others relying on black boxes. They all have the common goal to efficiently optimise quantum processes with the given resources. In general, using fewer parameters typically leads to a smaller search space and hence quicker convergence. However, the solutions are bad if the smaller search space does not include a good optimum. Changing the basis functions could be seen as an adjustment of the search space: With the right basis set, fewer parameters are required to find a good solution. Hence, the convergence is sped up.

This interpretation is supported by our results. In some cases, like the single qubit gate in Sec.~\ref{sec:qubit_gate}, we can clearly identify the characteristics of a good solution. The average converged pulses, as well as the average spectra, show specific properties that are more complex to produce with the Fourier and sinc bases than sigmoid or piecewise-constant. 
Similarly, broad spectra require many more sigmoid elements than Fourier or sincfunctions as shown on Sec.~\ref{sec:ising}. \textbf{Hence, a basis which reproduces known solutions to similar problems with few elements constitutes a promising choice.}
In other cases, like the Ising model problem described in Sec.~\ref{sec:ising}, no such recipe exists making the QOC application even more relevant. Still, in this example we can see that the ranking between the bases does not change with rising qubit number for the same problem class. Assuming that this transferability applies to other problems as well it allows for a cheap way to test for the best basis choice: \textbf{The best basis for a problem may be indicated by a test on a simplified version of the problem.}

\section{Conclusion}
\label{sec:conclusion}

The optimisation of different problems with different bases shows that the convergence period depends on the applied set of basis functions, with differences of up to a factor of 10 in the number of function evaluations required. For the presented examples, the ranking between the bases is dependent on the problem they are applied to. 
However, the ranking does not change with the problem complexity, see the number of qubits in the Ising model.

The convergence plots suggest that an educated choice of basis elements is likely to outperform the default, when attempting to reduce the number of basis elements. This is often the case for closed-loop optimisations, as most gradient-free optimisers are more effective in smaller search spaces and experimental evaluations are expensive. The right basis may lead to a faster and better solution with higher success probability. The basis choice should consider the decomposition of known solutions for similar problems or through trial-runs with a lower-complexity version of the problem (e.g., less qubits/energy levels or model-based open-loop).\\
In a situation where the required properties are well-known, they could be encoded directly into the optimisation for the cost of restricting the search space. A suggestion for such an approach for DRAG-criteria is given in the supplementary material~\footnote{see Supplementary Material V}.

\section{Outlook}
Thinking ahead, if one basis can outperform the other, what could their combination achieve? Especially in situations where no clear basis candidates present themselves, combining bases might be a valid strategy to explore. They could either be combined simultaneously, added in subsequent superiterations, or by creating hybrid bases, e.g., sincs with different bandwidths or frequency elements cut off by sigmoids. Our investigations only covered an exemplary set of bases, but others like the Slepian~\cite{lucarelli_quantum_2018}, Walsh~\cite{hayes_reducing_2011}\, windowed Gaussian~\cite{chalermpusitarak_frame-based_2021}, or B-spline~\cite{gunther_quantum_2021} basis might offer further advantages.
Furthermore, the presented results were produced with gradient-free dCRAB and partially confirmed with AD-based CRAB~\footnote{see Supplementary Material III}. We would be interested to see whether they hold as expected for other basis-flexible optimisation methods like GOAT and GROUP. 
In summary, we show how the basis choice affects the efficiency of optimisations and hope that our conclusions will inspire the use of a larger range of basis functions.

\emph{Code and data availability $-$} The code to perform the optimal control is available as QuOCS~\cite{Rossignolo2023}. The code for the analysis is available upon reasonable request. All the figures are available at~\cite{figshare_doi}. 

\begin{acknowledgments}
PR thanks Ian Yang, Robert Zeier, and Yuri Minoguchi for useful discussions. AP thanks Daniel Jaschke for useful discussions.
This project has received funding from the German Federal Ministry of
Education and Research (BMBF) under the funding program quantum
technologies $-$ from basic research to market $-$ with the grant QRydDemo.
We acknowledge financial support from the Italian Ministry of University
and Research (MUR) via the Department of
Excellence grant 2023-2027 Quantum Frontiers; from the European Union via
H2020 projects \mbox{EuRyQa}, \mbox{TEXTAROSSA}, and the Quantum Flagship project \mbox{Pasquans2}, the EU-QuantERA projects \mbox{QuantHEP} and T-NISQ, and
from the World Class Research Infrastructure $-$ Quantum Computing
and Simulation Center (QCSC) of Padova University, and the Italian National
Centre on HPC, Big Data and Quantum Computing.
The authors acknowledge support by the state of Baden-W{\"u}rttemberg through
bwHPC and the German Research Foundation (DFG) through grant no
INST 40/575-1 FUGG (JUSTUS 2 cluster). 
We acknowledge funding within the HPQC project by the Austrian Research Promotion 
Agency (FFG, project number 897481) supported by the European Union – NextGenerationEU.
We acknowledge HORIZON-CL4-2022-QUANTUM-01-SGA project under Grant 101113946 OpenSuperQ- Plus100, by Germany’s Excellence Strategy – Cluster of Excellence Matter and Light for Quantum Computing (ML4Q) EXC 2004/1 – 390534769, by the Helmholtz Validation Fund project “Qruise” (HVF-00096), and by the German Federal Ministry of Research (BMBF) under the project SPINNING (No. 13N16210), PASQUANS2.
\end{acknowledgments}

\bibliography{apssamp}

\begin{thebibliography}{48}%
\makeatletter
\providecommand \@ifxundefined [1]{%
 \@ifx{#1\undefined}
}%
\providecommand \@ifnum [1]{%
 \ifnum #1\expandafter \@firstoftwo
 \else \expandafter \@secondoftwo
 \fi
}%
\providecommand \@ifx [1]{%
 \ifx #1\expandafter \@firstoftwo
 \else \expandafter \@secondoftwo
 \fi
}%
\providecommand \natexlab [1]{#1}%
\providecommand \enquote  [1]{``#1''}%
\providecommand \bibnamefont  [1]{#1}%
\providecommand \bibfnamefont [1]{#1}%
\providecommand \citenamefont [1]{#1}%
\providecommand \href@noop [0]{\@secondoftwo}%
\providecommand \href [0]{\begingroup \@sanitize@url \@href}%
\providecommand \@href[1]{\@@startlink{#1}\@@href}%
\providecommand \@@href[1]{\endgroup#1\@@endlink}%
\providecommand \@sanitize@url [0]{\catcode `\\12\catcode `\$12\catcode `\&12\catcode `\#12\catcode `\^12\catcode `\_12\catcode `\%12\relax}%
\providecommand \@@startlink[1]{}%
\providecommand \@@endlink[0]{}%
\providecommand \url  [0]{\begingroup\@sanitize@url \@url }%
\providecommand \@url [1]{\endgroup\@href {#1}{\urlprefix }}%
\providecommand \urlprefix  [0]{URL }%
\providecommand \Eprint [0]{\href }%
\providecommand \doibase [0]{https://doi.org/}%
\providecommand \selectlanguage [0]{\@gobble}%
\providecommand \bibinfo  [0]{\@secondoftwo}%
\providecommand \bibfield  [0]{\@secondoftwo}%
\providecommand \translation [1]{[#1]}%
\providecommand \BibitemOpen [0]{}%
\providecommand \bibitemStop [0]{}%
\providecommand \bibitemNoStop [0]{.\EOS\space}%
\providecommand \EOS [0]{\spacefactor3000\relax}%
\providecommand \BibitemShut  [1]{\csname bibitem#1\endcsname}%
\let\auto@bib@innerbib\@empty
\bibitem [{\citenamefont {Koch}\ \emph {et~al.}(2022)\citenamefont {Koch}, \citenamefont {Boscain}, \citenamefont {Calarco}, \citenamefont {Dirr}, \citenamefont {Filipp}, \citenamefont {Glaser}, \citenamefont {Kosloff}, \citenamefont {Montangero}, \citenamefont {Schulte-Herbrüggen}, \citenamefont {Sugny},\ and\ \citenamefont {Wilhelm}}]{Koch2022}%
  \BibitemOpen
  \bibfield  {author} {\bibinfo {author} {\bibfnamefont {C.~P.}\ \bibnamefont {Koch}}, \bibinfo {author} {\bibfnamefont {U.}~\bibnamefont {Boscain}}, \bibinfo {author} {\bibfnamefont {T.}~\bibnamefont {Calarco}}, \bibinfo {author} {\bibfnamefont {G.}~\bibnamefont {Dirr}}, \bibinfo {author} {\bibfnamefont {S.}~\bibnamefont {Filipp}}, \bibinfo {author} {\bibfnamefont {S.~J.}\ \bibnamefont {Glaser}}, \bibinfo {author} {\bibfnamefont {R.}~\bibnamefont {Kosloff}}, \bibinfo {author} {\bibfnamefont {S.}~\bibnamefont {Montangero}}, \bibinfo {author} {\bibfnamefont {T.}~\bibnamefont {Schulte-Herbrüggen}}, \bibinfo {author} {\bibfnamefont {D.}~\bibnamefont {Sugny}},\ and\ \bibinfo {author} {\bibfnamefont {F.~K.}\ \bibnamefont {Wilhelm}},\ }\bibfield  {title} {\bibinfo {title} {Quantum optimal control in quantum technologies. strategic report on current status, visions and goals for research in europe},\ }\bibfield  {journal} {\bibinfo  {journal} {EPJ Quantum Technology}\ }\textbf {\bibinfo {volume} {9}},\ \href
  {https://doi.org/10.1140/epjqt/s40507-022-00138-x} {10.1140/epjqt/s40507-022-00138-x} (\bibinfo {year} {2022})\BibitemShut {NoStop}%
\bibitem [{\citenamefont {Glaser}\ \emph {et~al.}(2015)\citenamefont {Glaser}, \citenamefont {Boscain}, \citenamefont {Calarco}, \citenamefont {Koch}, \citenamefont {Köckenberger}, \citenamefont {Kosloff}, \citenamefont {Kuprov}, \citenamefont {Luy}, \citenamefont {Schirmer}, \citenamefont {Schulte-Herbrüggen}, \citenamefont {Sugny},\ and\ \citenamefont {Wilhelm}}]{Glaser2015}%
  \BibitemOpen
  \bibfield  {author} {\bibinfo {author} {\bibfnamefont {S.~J.}\ \bibnamefont {Glaser}}, \bibinfo {author} {\bibfnamefont {U.}~\bibnamefont {Boscain}}, \bibinfo {author} {\bibfnamefont {T.}~\bibnamefont {Calarco}}, \bibinfo {author} {\bibfnamefont {C.~P.}\ \bibnamefont {Koch}}, \bibinfo {author} {\bibfnamefont {W.}~\bibnamefont {Köckenberger}}, \bibinfo {author} {\bibfnamefont {R.}~\bibnamefont {Kosloff}}, \bibinfo {author} {\bibfnamefont {I.}~\bibnamefont {Kuprov}}, \bibinfo {author} {\bibfnamefont {B.}~\bibnamefont {Luy}}, \bibinfo {author} {\bibfnamefont {S.}~\bibnamefont {Schirmer}}, \bibinfo {author} {\bibfnamefont {T.}~\bibnamefont {Schulte-Herbrüggen}}, \bibinfo {author} {\bibfnamefont {D.}~\bibnamefont {Sugny}},\ and\ \bibinfo {author} {\bibfnamefont {F.~K.}\ \bibnamefont {Wilhelm}},\ }\bibfield  {title} {\bibinfo {title} {Training schrödinger’s cat: quantum optimal control: Strategic report on current status, visions and goals for research in europe},\ }\bibfield  {journal} {\bibinfo  {journal}
  {The European Physical Journal D}\ }\textbf {\bibinfo {volume} {69}},\ \href {https://doi.org/10.1140/epjd/e2015-60464-1} {10.1140/epjd/e2015-60464-1} (\bibinfo {year} {2015})\BibitemShut {NoStop}%
\bibitem [{\citenamefont {Brif}\ \emph {et~al.}(2010)\citenamefont {Brif}, \citenamefont {Chakrabarti},\ and\ \citenamefont {Rabitz}}]{Brif2010}%
  \BibitemOpen
  \bibfield  {author} {\bibinfo {author} {\bibfnamefont {C.}~\bibnamefont {Brif}}, \bibinfo {author} {\bibfnamefont {R.}~\bibnamefont {Chakrabarti}},\ and\ \bibinfo {author} {\bibfnamefont {H.}~\bibnamefont {Rabitz}},\ }\bibfield  {title} {\bibinfo {title} {Control of quantum phenomena: past, present and future},\ }\href {https://doi.org/10.1088/1367-2630/12/7/075008} {\bibfield  {journal} {\bibinfo  {journal} {New Journal of Physics}\ }\textbf {\bibinfo {volume} {12}},\ \bibinfo {pages} {075008} (\bibinfo {year} {2010})}\BibitemShut {NoStop}%
\bibitem [{\citenamefont {Rembold}\ \emph {et~al.}(2020)\citenamefont {Rembold}, \citenamefont {Oshnik}, \citenamefont {M{\"{u}}ller}, \citenamefont {Montangero}, \citenamefont {Calarco},\ and\ \citenamefont {Neu}}]{Rembold2020}%
  \BibitemOpen
  \bibfield  {author} {\bibinfo {author} {\bibfnamefont {P.}~\bibnamefont {Rembold}}, \bibinfo {author} {\bibfnamefont {N.}~\bibnamefont {Oshnik}}, \bibinfo {author} {\bibfnamefont {M.~M.}\ \bibnamefont {M{\"{u}}ller}}, \bibinfo {author} {\bibfnamefont {S.}~\bibnamefont {Montangero}}, \bibinfo {author} {\bibfnamefont {T.}~\bibnamefont {Calarco}},\ and\ \bibinfo {author} {\bibfnamefont {E.}~\bibnamefont {Neu}},\ }\bibfield  {title} {\bibinfo {title} {{Introduction to quantum optimal control for quantum sensing with nitrogen-vacancy centers in diamond}},\ }\href {https://doi.org/10.1116/5.0006785} {\bibfield  {journal} {\bibinfo  {journal} {AVS Quantum Science}\ }\textbf {\bibinfo {volume} {2}},\ \bibinfo {pages} {024701} (\bibinfo {year} {2020})}\BibitemShut {NoStop}%
\bibitem [{\citenamefont {Müller}\ \emph {et~al.}(2022{\natexlab{a}})\citenamefont {Müller}, \citenamefont {Said}, \citenamefont {Jelezko}, \citenamefont {Calarco},\ and\ \citenamefont {Montangero}}]{Muller2022}%
  \BibitemOpen
  \bibfield  {author} {\bibinfo {author} {\bibfnamefont {M.~M.}\ \bibnamefont {Müller}}, \bibinfo {author} {\bibfnamefont {R.~S.}\ \bibnamefont {Said}}, \bibinfo {author} {\bibfnamefont {F.}~\bibnamefont {Jelezko}}, \bibinfo {author} {\bibfnamefont {T.}~\bibnamefont {Calarco}},\ and\ \bibinfo {author} {\bibfnamefont {S.}~\bibnamefont {Montangero}},\ }\bibfield  {title} {\bibinfo {title} {One decade of quantum optimal control in the chopped random basis},\ }\href {https://doi.org/10.1088/1361-6633/ac723c} {\bibfield  {journal} {\bibinfo  {journal} {Reports on Progress in Physics}\ }\textbf {\bibinfo {volume} {85}},\ \bibinfo {pages} {076001} (\bibinfo {year} {2022}{\natexlab{a}})}\BibitemShut {NoStop}%
\bibitem [{\citenamefont {Omran}\ \emph {et~al.}(2019)\citenamefont {Omran}, \citenamefont {Levine}, \citenamefont {Keesling}, \citenamefont {Semeghini}, \citenamefont {Wang}, \citenamefont {Ebadi}, \citenamefont {Bernien}, \citenamefont {Zibrov}, \citenamefont {Pichler}, \citenamefont {Choi}, \citenamefont {Cui}, \citenamefont {Rossignolo}, \citenamefont {Rembold}, \citenamefont {Montangero}, \citenamefont {Calarco}, \citenamefont {Endres}, \citenamefont {Greiner}, \citenamefont {Vuletić},\ and\ \citenamefont {Lukin}}]{Omran2019}%
  \BibitemOpen
  \bibfield  {author} {\bibinfo {author} {\bibfnamefont {A.}~\bibnamefont {Omran}}, \bibinfo {author} {\bibfnamefont {H.}~\bibnamefont {Levine}}, \bibinfo {author} {\bibfnamefont {A.}~\bibnamefont {Keesling}}, \bibinfo {author} {\bibfnamefont {G.}~\bibnamefont {Semeghini}}, \bibinfo {author} {\bibfnamefont {T.~T.}\ \bibnamefont {Wang}}, \bibinfo {author} {\bibfnamefont {S.}~\bibnamefont {Ebadi}}, \bibinfo {author} {\bibfnamefont {H.}~\bibnamefont {Bernien}}, \bibinfo {author} {\bibfnamefont {A.~S.}\ \bibnamefont {Zibrov}}, \bibinfo {author} {\bibfnamefont {H.}~\bibnamefont {Pichler}}, \bibinfo {author} {\bibfnamefont {S.}~\bibnamefont {Choi}}, \bibinfo {author} {\bibfnamefont {J.}~\bibnamefont {Cui}}, \bibinfo {author} {\bibfnamefont {M.}~\bibnamefont {Rossignolo}}, \bibinfo {author} {\bibfnamefont {P.}~\bibnamefont {Rembold}}, \bibinfo {author} {\bibfnamefont {S.}~\bibnamefont {Montangero}}, \bibinfo {author} {\bibfnamefont {T.}~\bibnamefont {Calarco}}, \bibinfo {author} {\bibfnamefont {M.}~\bibnamefont
  {Endres}}, \bibinfo {author} {\bibfnamefont {M.}~\bibnamefont {Greiner}}, \bibinfo {author} {\bibfnamefont {V.}~\bibnamefont {Vuletić}},\ and\ \bibinfo {author} {\bibfnamefont {M.~D.}\ \bibnamefont {Lukin}},\ }\bibfield  {title} {\bibinfo {title} {Generation and manipulation of schrödinger cat states in rydberg atom arrays},\ }\href {https://doi.org/10.1126/science.aax9743} {\bibfield  {journal} {\bibinfo  {journal} {Science}\ }\textbf {\bibinfo {volume} {365}},\ \bibinfo {pages} {570–574} (\bibinfo {year} {2019})}\BibitemShut {NoStop}%
\bibitem [{\citenamefont {Pagano}\ \emph {et~al.}(2022)\citenamefont {Pagano}, \citenamefont {Weber}, \citenamefont {Jaschke}, \citenamefont {Pfau}, \citenamefont {Meinert}, \citenamefont {Montangero},\ and\ \citenamefont {Büchler}}]{Pagano2022}%
  \BibitemOpen
  \bibfield  {author} {\bibinfo {author} {\bibfnamefont {A.}~\bibnamefont {Pagano}}, \bibinfo {author} {\bibfnamefont {S.}~\bibnamefont {Weber}}, \bibinfo {author} {\bibfnamefont {D.}~\bibnamefont {Jaschke}}, \bibinfo {author} {\bibfnamefont {T.}~\bibnamefont {Pfau}}, \bibinfo {author} {\bibfnamefont {F.}~\bibnamefont {Meinert}}, \bibinfo {author} {\bibfnamefont {S.}~\bibnamefont {Montangero}},\ and\ \bibinfo {author} {\bibfnamefont {H.~P.}\ \bibnamefont {Büchler}},\ }\bibfield  {title} {\bibinfo {title} {Error budgeting for a controlled-phase gate with strontium-88 rydberg atoms},\ }\bibfield  {journal} {\bibinfo  {journal} {Physical Review Research}\ }\textbf {\bibinfo {volume} {4}},\ \href {https://doi.org/10.1103/physrevresearch.4.033019} {10.1103/physrevresearch.4.033019} (\bibinfo {year} {2022})\BibitemShut {NoStop}%
\bibitem [{\citenamefont {Rosi}\ \emph {et~al.}(2013)\citenamefont {Rosi}, \citenamefont {Bernard}, \citenamefont {Fabbri}, \citenamefont {Fallani}, \citenamefont {Fort}, \citenamefont {Inguscio}, \citenamefont {Calarco},\ and\ \citenamefont {Montangero}}]{Rosi2013}%
  \BibitemOpen
  \bibfield  {author} {\bibinfo {author} {\bibfnamefont {S.}~\bibnamefont {Rosi}}, \bibinfo {author} {\bibfnamefont {A.}~\bibnamefont {Bernard}}, \bibinfo {author} {\bibfnamefont {N.}~\bibnamefont {Fabbri}}, \bibinfo {author} {\bibfnamefont {L.}~\bibnamefont {Fallani}}, \bibinfo {author} {\bibfnamefont {C.}~\bibnamefont {Fort}}, \bibinfo {author} {\bibfnamefont {M.}~\bibnamefont {Inguscio}}, \bibinfo {author} {\bibfnamefont {T.}~\bibnamefont {Calarco}},\ and\ \bibinfo {author} {\bibfnamefont {S.}~\bibnamefont {Montangero}},\ }\bibfield  {title} {\bibinfo {title} {Fast closed-loop optimal control of ultracold atoms in an optical lattice},\ }\bibfield  {journal} {\bibinfo  {journal} {Physical Review A}\ }\textbf {\bibinfo {volume} {88}},\ \href {https://doi.org/10.1103/physreva.88.021601} {10.1103/physreva.88.021601} (\bibinfo {year} {2013})\BibitemShut {NoStop}%
\bibitem [{\citenamefont {van Frank}\ \emph {et~al.}(2016)\citenamefont {van Frank}, \citenamefont {Bonneau}, \citenamefont {Schmiedmayer}, \citenamefont {Hild}, \citenamefont {Gross}, \citenamefont {Cheneau}, \citenamefont {Bloch}, \citenamefont {Pichler}, \citenamefont {Negretti}, \citenamefont {Calarco},\ and\ \citenamefont {Montangero}}]{vanFrank2016}%
  \BibitemOpen
  \bibfield  {author} {\bibinfo {author} {\bibfnamefont {S.}~\bibnamefont {van Frank}}, \bibinfo {author} {\bibfnamefont {M.}~\bibnamefont {Bonneau}}, \bibinfo {author} {\bibfnamefont {J.}~\bibnamefont {Schmiedmayer}}, \bibinfo {author} {\bibfnamefont {S.}~\bibnamefont {Hild}}, \bibinfo {author} {\bibfnamefont {C.}~\bibnamefont {Gross}}, \bibinfo {author} {\bibfnamefont {M.}~\bibnamefont {Cheneau}}, \bibinfo {author} {\bibfnamefont {I.}~\bibnamefont {Bloch}}, \bibinfo {author} {\bibfnamefont {T.}~\bibnamefont {Pichler}}, \bibinfo {author} {\bibfnamefont {A.}~\bibnamefont {Negretti}}, \bibinfo {author} {\bibfnamefont {T.}~\bibnamefont {Calarco}},\ and\ \bibinfo {author} {\bibfnamefont {S.}~\bibnamefont {Montangero}},\ }\bibfield  {title} {\bibinfo {title} {Optimal control of complex atomic quantum systems},\ }\bibfield  {journal} {\bibinfo  {journal} {Scientific Reports}\ }\textbf {\bibinfo {volume} {6}},\ \href {https://doi.org/10.1038/srep34187} {10.1038/srep34187} (\bibinfo {year} {2016})\BibitemShut
  {NoStop}%
\bibitem [{\citenamefont {Heck}\ \emph {et~al.}(2018)\citenamefont {Heck}, \citenamefont {Vuculescu}, \citenamefont {Sørensen}, \citenamefont {Zoller}, \citenamefont {Andreasen}, \citenamefont {Bason}, \citenamefont {Ejlertsen}, \citenamefont {Elíasson}, \citenamefont {Haikka}, \citenamefont {Laustsen}, \citenamefont {Nielsen}, \citenamefont {Mao}, \citenamefont {Müller}, \citenamefont {Napolitano}, \citenamefont {Pedersen}, \citenamefont {Thorsen}, \citenamefont {Bergenholtz}, \citenamefont {Calarco}, \citenamefont {Montangero},\ and\ \citenamefont {Sherson}}]{Heck2018}%
  \BibitemOpen
  \bibfield  {author} {\bibinfo {author} {\bibfnamefont {R.}~\bibnamefont {Heck}}, \bibinfo {author} {\bibfnamefont {O.}~\bibnamefont {Vuculescu}}, \bibinfo {author} {\bibfnamefont {J.~J.}\ \bibnamefont {Sørensen}}, \bibinfo {author} {\bibfnamefont {J.}~\bibnamefont {Zoller}}, \bibinfo {author} {\bibfnamefont {M.~G.}\ \bibnamefont {Andreasen}}, \bibinfo {author} {\bibfnamefont {M.~G.}\ \bibnamefont {Bason}}, \bibinfo {author} {\bibfnamefont {P.}~\bibnamefont {Ejlertsen}}, \bibinfo {author} {\bibfnamefont {O.}~\bibnamefont {Elíasson}}, \bibinfo {author} {\bibfnamefont {P.}~\bibnamefont {Haikka}}, \bibinfo {author} {\bibfnamefont {J.~S.}\ \bibnamefont {Laustsen}}, \bibinfo {author} {\bibfnamefont {L.~L.}\ \bibnamefont {Nielsen}}, \bibinfo {author} {\bibfnamefont {A.}~\bibnamefont {Mao}}, \bibinfo {author} {\bibfnamefont {R.}~\bibnamefont {Müller}}, \bibinfo {author} {\bibfnamefont {M.}~\bibnamefont {Napolitano}}, \bibinfo {author} {\bibfnamefont {M.~K.}\ \bibnamefont {Pedersen}}, \bibinfo {author}
  {\bibfnamefont {A.~R.}\ \bibnamefont {Thorsen}}, \bibinfo {author} {\bibfnamefont {C.}~\bibnamefont {Bergenholtz}}, \bibinfo {author} {\bibfnamefont {T.}~\bibnamefont {Calarco}}, \bibinfo {author} {\bibfnamefont {S.}~\bibnamefont {Montangero}},\ and\ \bibinfo {author} {\bibfnamefont {J.~F.}\ \bibnamefont {Sherson}},\ }\bibfield  {title} {\bibinfo {title} {Remote optimization of an ultracold atoms experiment by experts and citizen scientists},\ }\bibfield  {journal} {\bibinfo  {journal} {Proceedings of the National Academy of Sciences}\ }\textbf {\bibinfo {volume} {115}},\ \href {https://doi.org/10.1073/pnas.1716869115} {10.1073/pnas.1716869115} (\bibinfo {year} {2018})\BibitemShut {NoStop}%
\bibitem [{\citenamefont {Cerfontaine}\ \emph {et~al.}(2020)\citenamefont {Cerfontaine}, \citenamefont {Botzem}, \citenamefont {Ritzmann}, \citenamefont {Humpohl}, \citenamefont {Ludwig}, \citenamefont {Schuh}, \citenamefont {Bougeard}, \citenamefont {Wieck},\ and\ \citenamefont {Bluhm}}]{Cerfontaine2020}%
  \BibitemOpen
  \bibfield  {author} {\bibinfo {author} {\bibfnamefont {P.}~\bibnamefont {Cerfontaine}}, \bibinfo {author} {\bibfnamefont {T.}~\bibnamefont {Botzem}}, \bibinfo {author} {\bibfnamefont {J.}~\bibnamefont {Ritzmann}}, \bibinfo {author} {\bibfnamefont {S.~S.}\ \bibnamefont {Humpohl}}, \bibinfo {author} {\bibfnamefont {A.}~\bibnamefont {Ludwig}}, \bibinfo {author} {\bibfnamefont {D.}~\bibnamefont {Schuh}}, \bibinfo {author} {\bibfnamefont {D.}~\bibnamefont {Bougeard}}, \bibinfo {author} {\bibfnamefont {A.~D.}\ \bibnamefont {Wieck}},\ and\ \bibinfo {author} {\bibfnamefont {H.}~\bibnamefont {Bluhm}},\ }\bibfield  {title} {\bibinfo {title} {Closed-loop control of a gaas-based singlet-triplet spin qubit with 99.5\% gate fidelity and low leakage},\ }\bibfield  {journal} {\bibinfo  {journal} {Nature Communications}\ }\textbf {\bibinfo {volume} {11}},\ \href {https://doi.org/10.1038/s41467-020-17865-3} {10.1038/s41467-020-17865-3} (\bibinfo {year} {2020})\BibitemShut {NoStop}%
\bibitem [{\citenamefont {Werninghaus}\ \emph {et~al.}(2021)\citenamefont {Werninghaus}, \citenamefont {Egger}, \citenamefont {Roy}, \citenamefont {Machnes}, \citenamefont {Wilhelm},\ and\ \citenamefont {Filipp}}]{werninghaus_leakage_2021}%
  \BibitemOpen
  \bibfield  {author} {\bibinfo {author} {\bibfnamefont {M.}~\bibnamefont {Werninghaus}}, \bibinfo {author} {\bibfnamefont {D.~J.}\ \bibnamefont {Egger}}, \bibinfo {author} {\bibfnamefont {F.}~\bibnamefont {Roy}}, \bibinfo {author} {\bibfnamefont {S.}~\bibnamefont {Machnes}}, \bibinfo {author} {\bibfnamefont {F.~K.}\ \bibnamefont {Wilhelm}},\ and\ \bibinfo {author} {\bibfnamefont {S.}~\bibnamefont {Filipp}},\ }\bibfield  {title} {\bibinfo {title} {Leakage reduction in fast superconducting qubit gates via optimal control},\ }\href {https://doi.org/10.1038/s41534-020-00346-2} {\bibfield  {journal} {\bibinfo  {journal} {npj Quantum Information}\ }\textbf {\bibinfo {volume} {7}},\ \bibinfo {pages} {1} (\bibinfo {year} {2021})},\ \bibinfo {note} {number: 1 Publisher: Nature Publishing Group}\BibitemShut {NoStop}%
\bibitem [{\citenamefont {Oshnik}\ \emph {et~al.}(2022)\citenamefont {Oshnik}, \citenamefont {Rembold}, \citenamefont {Calarco}, \citenamefont {Montangero}, \citenamefont {Neu},\ and\ \citenamefont {Müller}}]{oshnik_robust_2022}%
  \BibitemOpen
  \bibfield  {author} {\bibinfo {author} {\bibfnamefont {N.}~\bibnamefont {Oshnik}}, \bibinfo {author} {\bibfnamefont {P.}~\bibnamefont {Rembold}}, \bibinfo {author} {\bibfnamefont {T.}~\bibnamefont {Calarco}}, \bibinfo {author} {\bibfnamefont {S.}~\bibnamefont {Montangero}}, \bibinfo {author} {\bibfnamefont {E.}~\bibnamefont {Neu}},\ and\ \bibinfo {author} {\bibfnamefont {M.~M.}\ \bibnamefont {Müller}},\ }\bibfield  {title} {\bibinfo {title} {Robust magnetometry with single nitrogen-vacancy centers via two-step optimization},\ }\href {https://doi.org/10.1103/PhysRevA.106.013107} {\bibfield  {journal} {\bibinfo  {journal} {Physical Review A}\ }\textbf {\bibinfo {volume} {106}},\ \bibinfo {pages} {013107} (\bibinfo {year} {2022})},\ \bibinfo {note} {publisher: American Physical Society}\BibitemShut {NoStop}%
\bibitem [{\citenamefont {Vetter}\ \emph {et~al.}(2024)\citenamefont {Vetter}, \citenamefont {Reisser}, \citenamefont {Hirsch}, \citenamefont {Calarco}, \citenamefont {Motzoi}, \citenamefont {Jelezko},\ and\ \citenamefont {Müller}}]{Vetter2024}%
  \BibitemOpen
  \bibfield  {author} {\bibinfo {author} {\bibfnamefont {P.~J.}\ \bibnamefont {Vetter}}, \bibinfo {author} {\bibfnamefont {T.}~\bibnamefont {Reisser}}, \bibinfo {author} {\bibfnamefont {M.~G.}\ \bibnamefont {Hirsch}}, \bibinfo {author} {\bibfnamefont {T.}~\bibnamefont {Calarco}}, \bibinfo {author} {\bibfnamefont {F.}~\bibnamefont {Motzoi}}, \bibinfo {author} {\bibfnamefont {F.}~\bibnamefont {Jelezko}},\ and\ \bibinfo {author} {\bibfnamefont {M.~M.}\ \bibnamefont {Müller}},\ }\href@noop {} {\bibinfo {title} {Gate-set evaluation metrics for closed-loop optimal control on nitrogen-vacancy center ensembles in diamond}} (\bibinfo {year} {2024}),\ \Eprint {https://arxiv.org/abs/2403.00616} {arXiv:2403.00616 [quant-ph]} \BibitemShut {NoStop}%
\bibitem [{\citenamefont {Caneva}\ \emph {et~al.}(2011)\citenamefont {Caneva}, \citenamefont {Calarco},\ and\ \citenamefont {Montangero}}]{Caneva2011}%
  \BibitemOpen
  \bibfield  {author} {\bibinfo {author} {\bibfnamefont {T.}~\bibnamefont {Caneva}}, \bibinfo {author} {\bibfnamefont {T.}~\bibnamefont {Calarco}},\ and\ \bibinfo {author} {\bibfnamefont {S.}~\bibnamefont {Montangero}},\ }\bibfield  {title} {\bibinfo {title} {Chopped random-basis quantum optimization},\ }\href {https://doi.org/10.1103/PhysRevA.84.022326} {\bibfield  {journal} {\bibinfo  {journal} {Physical Review A - Atomic, Molecular, and Optical Physics}\ }\textbf {\bibinfo {volume} {84}},\ \bibinfo {pages} {22326} (\bibinfo {year} {2011})},\ \bibinfo {note} {arXiv: 1103.0855}\BibitemShut {NoStop}%
\bibitem [{\citenamefont {Rach}\ \emph {et~al.}(2015)\citenamefont {Rach}, \citenamefont {Müller}, \citenamefont {Calarco},\ and\ \citenamefont {Montangero}}]{Rach2015}%
  \BibitemOpen
  \bibfield  {author} {\bibinfo {author} {\bibfnamefont {N.}~\bibnamefont {Rach}}, \bibinfo {author} {\bibfnamefont {M.~M.}\ \bibnamefont {Müller}}, \bibinfo {author} {\bibfnamefont {T.}~\bibnamefont {Calarco}},\ and\ \bibinfo {author} {\bibfnamefont {S.}~\bibnamefont {Montangero}},\ }\bibfield  {title} {\bibinfo {title} {Dressing the chopped-random-basis optimization: A bandwidth-limited access to the trap-free landscape},\ }\bibfield  {journal} {\bibinfo  {journal} {Physical Review A}\ }\textbf {\bibinfo {volume} {92}},\ \href {https://doi.org/10.1103/physreva.92.062343} {10.1103/physreva.92.062343} (\bibinfo {year} {2015})\BibitemShut {NoStop}%
\bibitem [{\citenamefont {Machnes}\ \emph {et~al.}(2018)\citenamefont {Machnes}, \citenamefont {Assémat}, \citenamefont {Tannor},\ and\ \citenamefont {Wilhelm}}]{goat}%
  \BibitemOpen
  \bibfield  {author} {\bibinfo {author} {\bibfnamefont {S.}~\bibnamefont {Machnes}}, \bibinfo {author} {\bibfnamefont {E.}~\bibnamefont {Assémat}}, \bibinfo {author} {\bibfnamefont {D.}~\bibnamefont {Tannor}},\ and\ \bibinfo {author} {\bibfnamefont {F.~K.}\ \bibnamefont {Wilhelm}},\ }\bibfield  {title} {\bibinfo {title} {Tunable, {Flexible}, and {Efficient} {Optimization} of {Control} {Pulses} for {Practical} {Qubits}},\ }\href {https://doi.org/10.1103/PhysRevLett.120.150401} {\bibfield  {journal} {\bibinfo  {journal} {Physical Review Letters}\ }\textbf {\bibinfo {volume} {120}},\ \bibinfo {pages} {150401} (\bibinfo {year} {2018})}\BibitemShut {NoStop}%
\bibitem [{\citenamefont {Khaneja}\ \emph {et~al.}(2005)\citenamefont {Khaneja}, \citenamefont {Reiss}, \citenamefont {Kehlet}, \citenamefont {Schulte-Herbrüggen},\ and\ \citenamefont {Glaser}}]{KhanejaGRAPE}%
  \BibitemOpen
  \bibfield  {author} {\bibinfo {author} {\bibfnamefont {N.}~\bibnamefont {Khaneja}}, \bibinfo {author} {\bibfnamefont {T.}~\bibnamefont {Reiss}}, \bibinfo {author} {\bibfnamefont {C.}~\bibnamefont {Kehlet}}, \bibinfo {author} {\bibfnamefont {T.}~\bibnamefont {Schulte-Herbrüggen}},\ and\ \bibinfo {author} {\bibfnamefont {S.~J.}\ \bibnamefont {Glaser}},\ }\bibfield  {title} {\bibinfo {title} {{Optimal control of coupled spin dynamics: design of NMR pulse sequences by gradient ascent algorithms}},\ }\href {https://doi.org/https://doi.org/10.1016/j.jmr.2004.11.004} {\bibfield  {journal} {\bibinfo  {journal} {Journal of Magnetic Resonance}\ }\textbf {\bibinfo {volume} {172}},\ \bibinfo {pages} {296} (\bibinfo {year} {2005})}\BibitemShut {NoStop}%
\bibitem [{Note1()}]{Note1}%
  \BibitemOpen
  \bibinfo {note} {See Supplementary Material I}\BibitemShut {NoStop}%
\bibitem [{\citenamefont {Konnov}\ and\ \citenamefont {Krotov}(1999)}]{Konnov1999}%
  \BibitemOpen
  \bibfield  {author} {\bibinfo {author} {\bibfnamefont {A.~I.}\ \bibnamefont {Konnov}}\ and\ \bibinfo {author} {\bibfnamefont {V.~F.}\ \bibnamefont {Krotov}},\ }\bibfield  {title} {\bibinfo {title} {{On global methods for the successive improvement of control processes}},\ }\href {http://mi.mathnet.ru/at167} {\bibfield  {journal} {\bibinfo  {journal} {Avtomatika i Telemekhanika}\ }\textbf {\bibinfo {volume} {60}},\ \bibinfo {pages} {77} (\bibinfo {year} {1999})}\BibitemShut {NoStop}%
\bibitem [{\citenamefont {Sørensen}\ \emph {et~al.}(2018)\citenamefont {Sørensen}, \citenamefont {Aranburu}, \citenamefont {Heinzel},\ and\ \citenamefont {Sherson}}]{group}%
  \BibitemOpen
  \bibfield  {author} {\bibinfo {author} {\bibfnamefont {J.~J. W.~H.}\ \bibnamefont {Sørensen}}, \bibinfo {author} {\bibfnamefont {M.~O.}\ \bibnamefont {Aranburu}}, \bibinfo {author} {\bibfnamefont {T.}~\bibnamefont {Heinzel}},\ and\ \bibinfo {author} {\bibfnamefont {J.~F.}\ \bibnamefont {Sherson}},\ }\bibfield  {title} {\bibinfo {title} {Quantum optimal control in a chopped basis: {Applications} in control of {Bose}-{Einstein} condensates},\ }\href {https://doi.org/10.1103/PhysRevA.98.022119} {\bibfield  {journal} {\bibinfo  {journal} {Physical Review A}\ }\textbf {\bibinfo {volume} {98}},\ \bibinfo {pages} {022119} (\bibinfo {year} {2018})},\ \bibinfo {note} {publisher: American Physical Society}\BibitemShut {NoStop}%
\bibitem [{\citenamefont {Motzoi}\ \emph {et~al.}(2011)\citenamefont {Motzoi}, \citenamefont {Gambetta}, \citenamefont {Merkel},\ and\ \citenamefont {Wilhelm}}]{motzoi_optimal_2011}%
  \BibitemOpen
  \bibfield  {author} {\bibinfo {author} {\bibfnamefont {F.}~\bibnamefont {Motzoi}}, \bibinfo {author} {\bibfnamefont {J.~M.}\ \bibnamefont {Gambetta}}, \bibinfo {author} {\bibfnamefont {S.~T.}\ \bibnamefont {Merkel}},\ and\ \bibinfo {author} {\bibfnamefont {F.~K.}\ \bibnamefont {Wilhelm}},\ }\bibfield  {title} {\bibinfo {title} {Optimal control methods for rapidly time-varying {Hamiltonians}},\ }\href {https://doi.org/10.1103/PhysRevA.84.022307} {\bibfield  {journal} {\bibinfo  {journal} {Physical Review A}\ }\textbf {\bibinfo {volume} {84}},\ \bibinfo {pages} {022307} (\bibinfo {year} {2011})},\ \bibinfo {note} {publisher: American Physical Society}\BibitemShut {NoStop}%
\bibitem [{\citenamefont {Lloyd}\ and\ \citenamefont {Montangero}(2014)}]{Lloyd2014}%
  \BibitemOpen
  \bibfield  {author} {\bibinfo {author} {\bibfnamefont {S.}~\bibnamefont {Lloyd}}\ and\ \bibinfo {author} {\bibfnamefont {S.}~\bibnamefont {Montangero}},\ }\bibfield  {title} {\bibinfo {title} {Information theoretical analysis of quantum optimal control},\ }\href {https://doi.org/10.1103/PhysRevLett.113.010502} {\bibfield  {journal} {\bibinfo  {journal} {Physical Review Letters}\ }\textbf {\bibinfo {volume} {113}},\ \bibinfo {pages} {1} (\bibinfo {year} {2014})},\ \bibinfo {note} {arXiv: 1401.5047}\BibitemShut {NoStop}%
\bibitem [{\citenamefont {Caneva}\ \emph {et~al.}(2014)\citenamefont {Caneva}, \citenamefont {Silva}, \citenamefont {Fazio}, \citenamefont {Lloyd}, \citenamefont {Calarco},\ and\ \citenamefont {Montangero}}]{Caneva2014}%
  \BibitemOpen
  \bibfield  {author} {\bibinfo {author} {\bibfnamefont {T.}~\bibnamefont {Caneva}}, \bibinfo {author} {\bibfnamefont {A.}~\bibnamefont {Silva}}, \bibinfo {author} {\bibfnamefont {R.}~\bibnamefont {Fazio}}, \bibinfo {author} {\bibfnamefont {S.}~\bibnamefont {Lloyd}}, \bibinfo {author} {\bibfnamefont {T.}~\bibnamefont {Calarco}},\ and\ \bibinfo {author} {\bibfnamefont {S.}~\bibnamefont {Montangero}},\ }\bibfield  {title} {\bibinfo {title} {Complexity of controlling quantum many-body dynamics},\ }\bibfield  {journal} {\bibinfo  {journal} {Physical Review A}\ }\textbf {\bibinfo {volume} {89}},\ \href {https://doi.org/10.1103/physreva.89.042322} {10.1103/physreva.89.042322} (\bibinfo {year} {2014})\BibitemShut {NoStop}%
\bibitem [{\citenamefont {Müller}\ \emph {et~al.}(2022{\natexlab{b}})\citenamefont {Müller}, \citenamefont {Gherardini}, \citenamefont {Calarco}, \citenamefont {Montangero},\ and\ \citenamefont {Caruso}}]{Muller2022SR}%
  \BibitemOpen
  \bibfield  {author} {\bibinfo {author} {\bibfnamefont {M.~M.}\ \bibnamefont {Müller}}, \bibinfo {author} {\bibfnamefont {S.}~\bibnamefont {Gherardini}}, \bibinfo {author} {\bibfnamefont {T.}~\bibnamefont {Calarco}}, \bibinfo {author} {\bibfnamefont {S.}~\bibnamefont {Montangero}},\ and\ \bibinfo {author} {\bibfnamefont {F.}~\bibnamefont {Caruso}},\ }\bibfield  {title} {\bibinfo {title} {Information theoretical limits for quantum optimal control solutions: error scaling of noisy control channels},\ }\bibfield  {journal} {\bibinfo  {journal} {Scientific Reports}\ }\textbf {\bibinfo {volume} {12}},\ \href {https://doi.org/10.1038/s41598-022-25770-6} {10.1038/s41598-022-25770-6} (\bibinfo {year} {2022}{\natexlab{b}})\BibitemShut {NoStop}%
\bibitem [{\citenamefont {Jensen}\ \emph {et~al.}(2021)\citenamefont {Jensen}, \citenamefont {M\o{}ller}, \citenamefont {S\o{}rensen},\ and\ \citenamefont {Sherson}}]{Jensen2021}%
  \BibitemOpen
  \bibfield  {author} {\bibinfo {author} {\bibfnamefont {J.~H.~M.}\ \bibnamefont {Jensen}}, \bibinfo {author} {\bibfnamefont {F.~S.}\ \bibnamefont {M\o{}ller}}, \bibinfo {author} {\bibfnamefont {J.~J.}\ \bibnamefont {S\o{}rensen}},\ and\ \bibinfo {author} {\bibfnamefont {J.~F.}\ \bibnamefont {Sherson}},\ }\bibfield  {title} {\bibinfo {title} {Achieving fast high-fidelity optimal control of many-body quantum dynamics},\ }\href {https://doi.org/10.1103/PhysRevA.104.052210} {\bibfield  {journal} {\bibinfo  {journal} {Physical Review A}\ }\textbf {\bibinfo {volume} {104}},\ \bibinfo {pages} {052210} (\bibinfo {year} {2021})}\BibitemShut {NoStop}%
\bibitem [{\citenamefont {Li}(2023)}]{li_optimal_2023}%
  \BibitemOpen
  \bibfield  {author} {\bibinfo {author} {\bibfnamefont {X.}~\bibnamefont {Li}},\ }\bibfield  {title} {\bibinfo {title} {Optimal control of quantum state preparation and entanglement creation in two-qubit quantum system with bounded amplitude},\ }\href {https://doi.org/10.1038/s41598-023-41688-z} {\bibfield  {journal} {\bibinfo  {journal} {Scientific Reports}\ }\textbf {\bibinfo {volume} {13}},\ \bibinfo {pages} {14734} (\bibinfo {year} {2023})},\ \bibinfo {note} {publisher: Nature Publishing Group}\BibitemShut {NoStop}%
\bibitem [{\citenamefont {Chalermpusitarak}\ \emph {et~al.}(2021)\citenamefont {Chalermpusitarak}, \citenamefont {Tonekaboni}, \citenamefont {Wang}, \citenamefont {Norris}, \citenamefont {Viola},\ and\ \citenamefont {Paz-Silva}}]{chalermpusitarak_frame-based_2021}%
  \BibitemOpen
  \bibfield  {author} {\bibinfo {author} {\bibfnamefont {T.}~\bibnamefont {Chalermpusitarak}}, \bibinfo {author} {\bibfnamefont {B.}~\bibnamefont {Tonekaboni}}, \bibinfo {author} {\bibfnamefont {Y.}~\bibnamefont {Wang}}, \bibinfo {author} {\bibfnamefont {L.~M.}\ \bibnamefont {Norris}}, \bibinfo {author} {\bibfnamefont {L.}~\bibnamefont {Viola}},\ and\ \bibinfo {author} {\bibfnamefont {G.~A.}\ \bibnamefont {Paz-Silva}},\ }\bibfield  {title} {\bibinfo {title} {Frame-{Based} {Filter}-{Function} {Formalism} for {Quantum} {Characterization} and {Control}},\ }\href {https://doi.org/10.1103/PRXQuantum.2.030315} {\bibfield  {journal} {\bibinfo  {journal} {PRX Quantum}\ }\textbf {\bibinfo {volume} {2}},\ \bibinfo {pages} {030315} (\bibinfo {year} {2021})}\BibitemShut {NoStop}%
\bibitem [{\citenamefont {Motzoi}\ \emph {et~al.}(2009)\citenamefont {Motzoi}, \citenamefont {Gambetta}, \citenamefont {Rebentrost},\ and\ \citenamefont {Wilhelm}}]{Motzoi2009}%
  \BibitemOpen
  \bibfield  {author} {\bibinfo {author} {\bibfnamefont {F.}~\bibnamefont {Motzoi}}, \bibinfo {author} {\bibfnamefont {J.~M.}\ \bibnamefont {Gambetta}}, \bibinfo {author} {\bibfnamefont {P.}~\bibnamefont {Rebentrost}},\ and\ \bibinfo {author} {\bibfnamefont {F.~K.}\ \bibnamefont {Wilhelm}},\ }\bibfield  {title} {\bibinfo {title} {Simple {Pulses} for {Elimination} of {Leakage} in {Weakly} {Nonlinear} {Qubits}},\ }\href {https://doi.org/10.1103/PhysRevLett.103.110501} {\bibfield  {journal} {\bibinfo  {journal} {Phys. Rev. Lett.}\ }\textbf {\bibinfo {volume} {103}},\ \bibinfo {pages} {110501} (\bibinfo {year} {2009})}\BibitemShut {NoStop}%
\bibitem [{Note2()}]{Note2}%
  \BibitemOpen
  \bibinfo {note} {See Supplementary Material III}\BibitemShut {NoStop}%
\bibitem [{\citenamefont {Rossignolo}\ \emph {et~al.}(2023)\citenamefont {Rossignolo}, \citenamefont {Reisser}, \citenamefont {Marshall}, \citenamefont {Rembold}, \citenamefont {Pagano}, \citenamefont {Vetter}, \citenamefont {Said}, \citenamefont {Müller}, \citenamefont {Motzoi}, \citenamefont {Calarco}, \citenamefont {Jelezko},\ and\ \citenamefont {Montangero}}]{Rossignolo2023}%
  \BibitemOpen
  \bibfield  {author} {\bibinfo {author} {\bibfnamefont {M.}~\bibnamefont {Rossignolo}}, \bibinfo {author} {\bibfnamefont {T.}~\bibnamefont {Reisser}}, \bibinfo {author} {\bibfnamefont {A.}~\bibnamefont {Marshall}}, \bibinfo {author} {\bibfnamefont {P.}~\bibnamefont {Rembold}}, \bibinfo {author} {\bibfnamefont {A.}~\bibnamefont {Pagano}}, \bibinfo {author} {\bibfnamefont {P.~J.}\ \bibnamefont {Vetter}}, \bibinfo {author} {\bibfnamefont {R.~S.}\ \bibnamefont {Said}}, \bibinfo {author} {\bibfnamefont {M.~M.}\ \bibnamefont {Müller}}, \bibinfo {author} {\bibfnamefont {F.}~\bibnamefont {Motzoi}}, \bibinfo {author} {\bibfnamefont {T.}~\bibnamefont {Calarco}}, \bibinfo {author} {\bibfnamefont {F.}~\bibnamefont {Jelezko}},\ and\ \bibinfo {author} {\bibfnamefont {S.}~\bibnamefont {Montangero}},\ }\bibfield  {title} {\bibinfo {title} {{QuOCS: The Quantum Optimal Control Suite}},\ }\href {https://doi.org/https://doi.org/10.1016/j.cpc.2023.108782} {\bibfield  {journal} {\bibinfo  {journal} {{Computer Physics
  Communications}}\ ,\ \bibinfo {pages} {108782}} (\bibinfo {year} {2023})}\BibitemShut {NoStop}%
\bibitem [{Note3()}]{Note3}%
  \BibitemOpen
  \bibinfo {note} {See Supplementary Material I}\BibitemShut {NoStop}%
\bibitem [{\citenamefont {Rembold}(2022)}]{rembold_quantum_2022}%
  \BibitemOpen
  \bibfield  {author} {\bibinfo {author} {\bibfnamefont {P.}~\bibnamefont {Rembold}},\ }\emph {\bibinfo {title} {Quantum {Optimal} {Control} of {Spin} {Systems} and {Trapped} {Atoms}}},\ \href {https://kups.ub.uni-koeln.de/61563/} {\bibinfo {type} {{PhD} thesis}},\ \bibinfo  {school} {Universität zu Köln and Università degli Studi di Padova} (\bibinfo {year} {2022})\BibitemShut {NoStop}%
\bibitem [{Note4()}]{Note4}%
  \BibitemOpen
  \bibinfo {note} {See Supplementary Material II}\BibitemShut {NoStop}%
\bibitem [{\citenamefont {Doherty}\ \emph {et~al.}(2013)\citenamefont {Doherty}, \citenamefont {Manson}, \citenamefont {Delaney}, \citenamefont {Jelezko}, \citenamefont {Wrachtrup},\ and\ \citenamefont {Hollenberg}}]{doherty_nitrogen-vacancy_2013}%
  \BibitemOpen
  \bibfield  {author} {\bibinfo {author} {\bibfnamefont {M.~W.}\ \bibnamefont {Doherty}}, \bibinfo {author} {\bibfnamefont {N.~B.}\ \bibnamefont {Manson}}, \bibinfo {author} {\bibfnamefont {P.}~\bibnamefont {Delaney}}, \bibinfo {author} {\bibfnamefont {F.}~\bibnamefont {Jelezko}}, \bibinfo {author} {\bibfnamefont {J.}~\bibnamefont {Wrachtrup}},\ and\ \bibinfo {author} {\bibfnamefont {L.~C.~L.}\ \bibnamefont {Hollenberg}},\ }\bibfield  {title} {\bibinfo {title} {The nitrogen-vacancy colour centre in diamond},\ }\href {https://doi.org/10.1016/j.physrep.2013.02.001} {\bibfield  {journal} {\bibinfo  {journal} {Physics Reports}\ }\bibinfo {series} {The nitrogen-vacancy colour centre in diamond},\ \textbf {\bibinfo {volume} {528}},\ \bibinfo {pages} {1} (\bibinfo {year} {2013})}\BibitemShut {NoStop}%
\bibitem [{\citenamefont {Vetter}\ \emph {et~al.}(2022)\citenamefont {Vetter}, \citenamefont {Marshall}, \citenamefont {Genov}, \citenamefont {Weiss}, \citenamefont {Striegler}, \citenamefont {Gro\ss{}mann}, \citenamefont {Oviedo-Casado}, \citenamefont {Cerrillo}, \citenamefont {Prior}, \citenamefont {Neumann},\ and\ \citenamefont {Jelezko}}]{Vetter2022}%
  \BibitemOpen
  \bibfield  {author} {\bibinfo {author} {\bibfnamefont {P.~J.}\ \bibnamefont {Vetter}}, \bibinfo {author} {\bibfnamefont {A.}~\bibnamefont {Marshall}}, \bibinfo {author} {\bibfnamefont {G.~T.}\ \bibnamefont {Genov}}, \bibinfo {author} {\bibfnamefont {T.~F.}\ \bibnamefont {Weiss}}, \bibinfo {author} {\bibfnamefont {N.}~\bibnamefont {Striegler}}, \bibinfo {author} {\bibfnamefont {E.~F.}\ \bibnamefont {Gro\ss{}mann}}, \bibinfo {author} {\bibfnamefont {S.}~\bibnamefont {Oviedo-Casado}}, \bibinfo {author} {\bibfnamefont {J.}~\bibnamefont {Cerrillo}}, \bibinfo {author} {\bibfnamefont {J.}~\bibnamefont {Prior}}, \bibinfo {author} {\bibfnamefont {P.}~\bibnamefont {Neumann}},\ and\ \bibinfo {author} {\bibfnamefont {F.}~\bibnamefont {Jelezko}},\ }\bibfield  {title} {\bibinfo {title} {Zero- and low-field sensing with nitrogen-vacancy centers},\ }\href {https://doi.org/10.1103/PhysRevApplied.17.044028} {\bibfield  {journal} {\bibinfo  {journal} {Phys. Rev. Appl.}\ }\textbf {\bibinfo {volume} {17}},\ \bibinfo {pages}
  {044028} (\bibinfo {year} {2022})}\BibitemShut {NoStop}%
\bibitem [{\citenamefont {Krantz}\ \emph {et~al.}(2019)\citenamefont {Krantz}, \citenamefont {Kjaergaard}, \citenamefont {Yan}, \citenamefont {Orlando}, \citenamefont {Gustavsson},\ and\ \citenamefont {Oliver}}]{krantz_quantum_2019}%
  \BibitemOpen
  \bibfield  {author} {\bibinfo {author} {\bibfnamefont {P.}~\bibnamefont {Krantz}}, \bibinfo {author} {\bibfnamefont {M.}~\bibnamefont {Kjaergaard}}, \bibinfo {author} {\bibfnamefont {F.}~\bibnamefont {Yan}}, \bibinfo {author} {\bibfnamefont {T.~P.}\ \bibnamefont {Orlando}}, \bibinfo {author} {\bibfnamefont {S.}~\bibnamefont {Gustavsson}},\ and\ \bibinfo {author} {\bibfnamefont {W.~D.}\ \bibnamefont {Oliver}},\ }\bibfield  {title} {\bibinfo {title} {A quantum engineer's guide to superconducting qubits},\ }\bibfield  {journal} {\bibinfo  {journal} {Applied Physics Reviews}\ }\textbf {\bibinfo {volume} {6}},\ \href {https://doi.org/10.1063/1.5089550} {10.1063/1.5089550} (\bibinfo {year} {2019}),\ \bibinfo {note} {arXiv: 1904.06560 Publisher: AIP Publishing LLC}\BibitemShut {NoStop}%
\bibitem [{\citenamefont {Caneva}\ \emph {et~al.}(2009)\citenamefont {Caneva}, \citenamefont {Murphy}, \citenamefont {Calarco}, \citenamefont {Fazio}, \citenamefont {Montangero}, \citenamefont {Giovannetti},\ and\ \citenamefont {Santoro}}]{caneva_optimal_2009}%
  \BibitemOpen
  \bibfield  {author} {\bibinfo {author} {\bibfnamefont {T.}~\bibnamefont {Caneva}}, \bibinfo {author} {\bibfnamefont {M.}~\bibnamefont {Murphy}}, \bibinfo {author} {\bibfnamefont {T.}~\bibnamefont {Calarco}}, \bibinfo {author} {\bibfnamefont {R.}~\bibnamefont {Fazio}}, \bibinfo {author} {\bibfnamefont {S.}~\bibnamefont {Montangero}}, \bibinfo {author} {\bibfnamefont {V.}~\bibnamefont {Giovannetti}},\ and\ \bibinfo {author} {\bibfnamefont {G.~E.}\ \bibnamefont {Santoro}},\ }\bibfield  {title} {\bibinfo {title} {Optimal {Control} at the {Quantum} {Speed} {Limit}},\ }\href {https://doi.org/10.1103/PhysRevLett.103.240501} {\bibfield  {journal} {\bibinfo  {journal} {Physical Review Letters}\ }\textbf {\bibinfo {volume} {103}},\ \bibinfo {pages} {240501} (\bibinfo {year} {2009})}\BibitemShut {NoStop}%
\bibitem [{\citenamefont {Safaei}\ \emph {et~al.}(2009)\citenamefont {Safaei}, \citenamefont {Montangero}, \citenamefont {Taddei},\ and\ \citenamefont {Fazio}}]{Safaei2009}%
  \BibitemOpen
  \bibfield  {author} {\bibinfo {author} {\bibfnamefont {S.}~\bibnamefont {Safaei}}, \bibinfo {author} {\bibfnamefont {S.}~\bibnamefont {Montangero}}, \bibinfo {author} {\bibfnamefont {F.}~\bibnamefont {Taddei}},\ and\ \bibinfo {author} {\bibfnamefont {R.}~\bibnamefont {Fazio}},\ }\bibfield  {title} {\bibinfo {title} {Optimized single-qubit gates for josephson phase qubits},\ }\bibfield  {journal} {\bibinfo  {journal} {Physical Review B}\ }\textbf {\bibinfo {volume} {79}},\ \href {https://doi.org/10.1103/physrevb.79.064524} {10.1103/physrevb.79.064524} (\bibinfo {year} {2009})\BibitemShut {NoStop}%
\bibitem [{Note5()}]{Note5}%
  \BibitemOpen
  \bibinfo {note} {See Supplementary Material V}\BibitemShut {NoStop}%
\bibitem [{Note6()}]{Note6}%
  \BibitemOpen
  \bibinfo {note} {See Supplementary Material I}\BibitemShut {NoStop}%
\bibitem [{Note7()}]{Note7}%
  \BibitemOpen
  \bibinfo {note} {See Supplementary Material V}\BibitemShut {NoStop}%
\bibitem [{Note8()}]{Note8}%
  \BibitemOpen
  \bibinfo {note} {See Supplementary Material V}\BibitemShut {NoStop}%
\bibitem [{\citenamefont {Lucarelli}(2018)}]{lucarelli_quantum_2018}%
  \BibitemOpen
  \bibfield  {author} {\bibinfo {author} {\bibfnamefont {D.}~\bibnamefont {Lucarelli}},\ }\bibfield  {title} {\bibinfo {title} {Quantum optimal control via gradient ascent in function space and the time-bandwidth quantum speed limit},\ }\href {https://doi.org/10.1103/PhysRevA.97.062346} {\bibfield  {journal} {\bibinfo  {journal} {Physical Review A}\ }\textbf {\bibinfo {volume} {97}},\ \bibinfo {pages} {062346} (\bibinfo {year} {2018})}\BibitemShut {NoStop}%
\bibitem [{\citenamefont {Hayes}\ \emph {et~al.}(2011)\citenamefont {Hayes}, \citenamefont {Khodjasteh}, \citenamefont {Viola},\ and\ \citenamefont {Biercuk}}]{hayes_reducing_2011}%
  \BibitemOpen
  \bibfield  {author} {\bibinfo {author} {\bibfnamefont {D.}~\bibnamefont {Hayes}}, \bibinfo {author} {\bibfnamefont {K.}~\bibnamefont {Khodjasteh}}, \bibinfo {author} {\bibfnamefont {L.}~\bibnamefont {Viola}},\ and\ \bibinfo {author} {\bibfnamefont {M.~J.}\ \bibnamefont {Biercuk}},\ }\bibfield  {title} {\bibinfo {title} {Reducing sequencing complexity in dynamical quantum error suppression by {Walsh} modulation},\ }\href {https://doi.org/10.1103/PhysRevA.84.062323} {\bibfield  {journal} {\bibinfo  {journal} {Physical Review A}\ }\textbf {\bibinfo {volume} {84}},\ \bibinfo {pages} {062323} (\bibinfo {year} {2011})},\ \bibinfo {note} {publisher: American Physical Society}\BibitemShut {NoStop}%
\bibitem [{\citenamefont {Günther}\ \emph {et~al.}(2021)\citenamefont {Günther}, \citenamefont {Petersson},\ and\ \citenamefont {DuBois}}]{gunther_quantum_2021}%
  \BibitemOpen
  \bibfield  {author} {\bibinfo {author} {\bibfnamefont {S.}~\bibnamefont {Günther}}, \bibinfo {author} {\bibfnamefont {N.~A.}\ \bibnamefont {Petersson}},\ and\ \bibinfo {author} {\bibfnamefont {J.~L.}\ \bibnamefont {DuBois}},\ }\bibfield  {title} {\bibinfo {title} {Quantum optimal control for pure-state preparation using one initial state},\ }\href {https://doi.org/10.1116/5.0060262} {\bibfield  {journal} {\bibinfo  {journal} {AVS Quantum Science}\ }\textbf {\bibinfo {volume} {3}},\ \bibinfo {pages} {043801} (\bibinfo {year} {2021})}\BibitemShut {NoStop}%
\bibitem [{Note9()}]{Note9}%
  \BibitemOpen
  \bibinfo {note} {See Supplementary Material III}\BibitemShut {NoStop}%
\bibitem [{\citenamefont {Pagano}\ \emph {et~al.}(2024)\citenamefont {Pagano}, \citenamefont {Müller}, \citenamefont {Calarco}, \citenamefont {Montangero},\ and\ \citenamefont {Rembold}}]{figshare_doi}%
  \BibitemOpen
  \bibfield  {author} {\bibinfo {author} {\bibfnamefont {A.}~\bibnamefont {Pagano}}, \bibinfo {author} {\bibfnamefont {M.~M.}\ \bibnamefont {Müller}}, \bibinfo {author} {\bibfnamefont {T.}~\bibnamefont {Calarco}}, \bibinfo {author} {\bibfnamefont {S.}~\bibnamefont {Montangero}},\ and\ \bibinfo {author} {\bibfnamefont {P.}~\bibnamefont {Rembold}},\ }\bibfield  {title} {\bibinfo {title} {Figures and supplementary material of {T}he {R}ole of {B}ases in {Q}uantum {O}ptimal {C}ontrol},\ }\bibfield  {journal} {\bibinfo  {journal} {Figshare}\ }\href {https://doi.org/10.6084/m9.figshare.c.7277647.v2} {10.6084/m9.figshare.c.7277647.v2} (\bibinfo {year} {2024})\BibitemShut {NoStop}%
\end{thebibliography}%

\newpage
\thispagestyle{empty}
\mbox{}
\newpage
\newpage
\thispagestyle{empty}
\mbox{}
\includepdf[pages={1}]{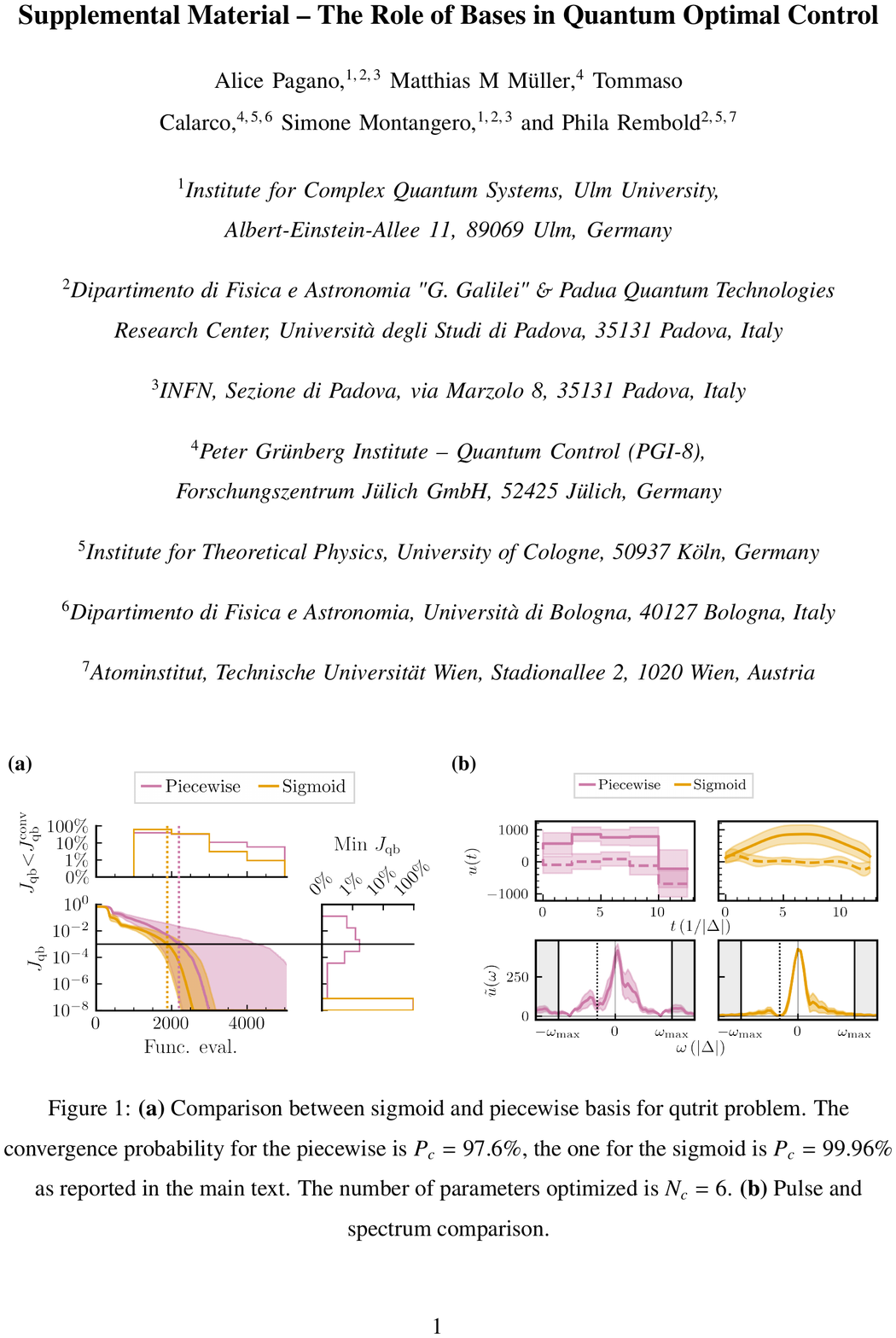}
\newpage
\thispagestyle{empty}
\mbox{}
\includepdf[pages={2}]{role_of_QOC_supps.pdf}
\newpage
\thispagestyle{empty}
\mbox{}
\includepdf[pages={3}]{role_of_QOC_supps.pdf}
\newpage
\thispagestyle{empty}
\mbox{}
\includepdf[pages={4}]{role_of_QOC_supps.pdf}
\newpage
\thispagestyle{empty}
\mbox{}
\includepdf[pages={5}]{role_of_QOC_supps.pdf}
\newpage
\thispagestyle{empty}
\mbox{}
\includepdf[pages={6}]{role_of_QOC_supps.pdf}
\newpage
\thispagestyle{empty}
\mbox{}
\includepdf[pages={7}]{role_of_QOC_supps.pdf}
\newpage
\thispagestyle{empty}
\mbox{}
\includepdf[pages={8}]{role_of_QOC_supps.pdf}
\newpage
\thispagestyle{empty}
\mbox{}
\includepdf[pages={9}]{role_of_QOC_supps.pdf}
\newpage
\thispagestyle{empty}
\mbox{}
\includepdf[pages={10}]{role_of_QOC_supps.pdf}
\end{document}